\documentclass[12pt]{article}
\usepackage{amsmath,amssymb}
\usepackage[left=2.9cm,right=2.9cm]{geometry}
\usepackage{color}
\usepackage[colorlinks=true
,urlcolor=blue
,anchorcolor=blue
,citecolor=blue
,filecolor=blue
,linkcolor=blue
,menucolor=blue
,linktocpage=true
,pdfproducer=medialab
,pdfa=true]{hyperref}
\usepackage{ifthen}
\newboolean{eps}
\setboolean{eps}{false}
\makeatletter
\let\over\@@over
\makeatother

\newcommand{\bea}{\begin{eqnarray}}
\newcommand{\eea}{\end{eqnarray}}
\newcommand{\be}{\begin{equation}}
\newcommand{\ee}{\end{equation}}
\newcommand{\ii}{\mathrm{i}}
\newcommand{\dd}{\mathrm{d}}
\newcommand{\e}[1]{\mathrm{e}^{#1}}
\DeclareMathOperator{\sech}{sech}
\DeclareMathOperator{\End}{End}
\DeclareMathOperator{\Hom}{Hom}
\DeclareMathOperator{\pr}{pr}
\DeclareMathOperator{\mmod}{mod}

\def\bar{\overline}
\def\b{\bar}

\numberwithin{equation}{section}

\begin{document}
\begin{flushright} \small
UUITP-29/24
 \end{flushright}
\smallskip
\begin{center} \Large
{\bf The Equivariant $B$ model}
 ~\\[12mm] \normalsize
{\bf Guido Festuccia$^a$, Roman Mauch$^a$, Maxim Zabzine$^{a,b}$}~ \\[8mm]
 {\small\it
  $^a$Department of Physics and Astronomy, Uppsala University,\\Box 516, SE-75120 Uppsala, Sweden\\  
  \vspace{.5cm}
  $^b$Centre for Geometry and Physics, Uppsala University,\\
  Box 480, SE-75106 Uppsala, Sweden}
\end{center}
\vspace{7mm}
\begin{abstract}

\noindent 
In this work, we introduce an equivariant deformation of the $B$ model on the sphere with a $U(1)$-action. We present the deformed supersymmetry transformations and corresponding Lagrangians and study observables in the supercharge cohomology. The inclusion of equivariance allows for the introduction of novel, position-dependent observables on the sphere, which have no counterparts in the conventional 
$B$ model. Two specific cases we explore in detail are position-dependent superpotentials and complex structure deformations.  In both instances, the theory exhibits notable differences from the standard $B$ model, revealing intriguing new features.

\end{abstract}

\eject
\normalsize

\tableofcontents

\section{Introduction}

The topological $B$ model was originally defined through the topological twist of an $\mathcal{N}=(2,2)$ supersymmetric non-linear sigma model \cite{Witten:1991zz,Labastida:1991qq}. The $B$ model can be promoted to
a type of topological string theory defined over an arbitrary Riemann surface $\Sigma_g$ and it depends on the complex structure of the target space, a Calabi-Yau manifold. The physical observables are related to holomorphic quantities, and its correlation functions capture the geometry of the moduli space of complex structures. The $B$ model is central in mirror symmetry, where it is dual to the A-model, which depends on the Kähler structure (for a review see \cite{Hori:2003ic}). 

 In this work we study the equivariant deformation of the $B$ model with respect to a $U(1)$-action on $\Sigma_g$. This restricts the possible compact $\Sigma_g$ we can consider to either $S^2$ or $T^2$. One can also study a version of the equivariant $B$ model on $\mathbb{C}$, the so called $\Omega$-background. In this work we focus on $S^2$ where the $U(1)$ action has fixed points. The $\Omega$-background can then be related to the theory on a hemisphere. 
The equivariant deformation of the $B$ model transformations was originally introduced in \cite{Yagi:2014toa} and \cite{Nekrasov:2018pqq}  which, however, did not explore extensively the resulting theory. 

In this work we present a systematic study of the equivariant $B$ model. We clarify its relation to a ${\cal N}=(2,2)$ non-linear sigma model for twisted chiral multiplets coupled to a specific supergravity background as studied in~\cite{Closset:2014pda,Closset:2015rna}.  
We explore a new class of observables that may explicitly depend on the worldsheet coordinates, and compute the corresponding correlators via the localization technique. We focus on two cases: the equivariant extension of the topological Landau-Ginzburg model and the equivariant extension of target-space complex structure deformations. 

The $B$ model can be coupled to a superpotential 
$W$, a setup known as the topological Landau-Ginzburg (LG) model \cite{Vafa:1990mu}. We consider the equivariant extension of this model, which allows for a superpotential $W(\phi,\vartheta)$ that varies with the latitude $\vartheta$ between the two poles of $S^2$, with different chiral rings associated with the north and south poles. The localization locus in this setup is given by non-trivial maps to the target space. The structures which appear here are very much reminiscent of topological anti-topological fusion  \cite{Cecotti:1991me} but in a purely holomorphic setting.
We also explore complex structure deformations $\mu^i{}_{\bar{j}} (\phi, \bar{\phi}, \vartheta)$ with explicit worldsheet dependence. These can be considered as an equivariant $B$ model on a family of Calabi-Yau manifolds in analogy with the picture 
suggested in \cite{Witten:1988xj}. 

The paper is organized as follows: In Section  \ref{sec:Bmodel} we review the standard $B$ model and establish the conventions for the subsequent discussion. In Section \ref{sec:equivB} we introduce the equivariant deformation of the $B$ model, including the deformed supersymmetry algebra, Lagrangians, and study its observables. In Section \ref{sec:superpotentials} we examine the case where the superpotential $W(\phi,\vartheta)$ depends explicitly on the worldsheet coordinate. Section \ref{sec:deformations} addresses complex structure deformations that also explicitly depend on the worldsheet coordinate. Finally, in Section \ref{sec:summary} we provide a summary of the paper and discuss open problems. The paper is supplemented by five appendices, in which we collect various technical details.

\section{Review of \texorpdfstring{$B$}{B} model}
\label{sec:Bmodel}

In this section we review the field content, supersymmetry transformations, Lagrangians and observables of the topological $B$ model~\cite{Witten:1991zz,Labastida:1991qq}. In order to better fit with the discussion of the equivariant $B$ model, our setup differs in some respects from the literature. We comment on these differences and motivate our choices.

\subsection{Supersymmetry and Lagrangians}
Let $\Phi:\Sigma_g\rightarrow X$ be a map from a two-dimensional Riemann surface $\Sigma_g$ to a K\"ahler target space\footnote{To define the fermion determinant $X$ needs to be Calabi-Yau unless $\Sigma_g$ is $T^2$~\cite{Witten:1991zz}.} $X$.
The only nonzero Christoffel symbols for the K\"ahler metric $g_{i\bar j}=\partial_i\partial_{\bar j} K$ on $X$ are:
\begin{equation}
\Gamma^{i}_{j k}=g^{i\bar l} \partial_{j} g_{k \bar l}=g^{i\bar l} \partial_{k} g_{j \bar l}~,\quad \Gamma^{\bar i}_{\bar j \bar k}=g^{l\bar i} \partial_{\bar j} g_{l \bar k}=g^{l\bar i} \partial_{\bar k } g_{l \bar j}~.
\end{equation}
In terms of these, up to symmetry, the nonzero components of the Riemann tensor are 
\bea
{R^{i}}_{j \bar l k}=\partial_{\bar l} \Gamma^{i}_{k j}~,\quad {R^{\bar i}}_{\bar j  l \bar k}=\partial_{l} \Gamma^{\bar i}_{\bar k \bar j}~.
\eea
  The field content of the $B$ model is given by 
  \begin{itemize}
\item the map $\Phi$ expressed in local holomorphic coordinates $\phi^i(z,\bar{z}), \phi^{\bar i}(z,\bar{z})$~, 
 \item  fermionic fields $ \eta^{\bar i},\theta^{\bar i}\in\Phi^\ast(T^{0,1}X)$   and $\rho^i\in\Omega^1(\Sigma)\otimes\Phi^\ast(T^{1,0}X)$~,
 \item   auxiliary fields $ \beta^{\bar i}\in\Phi^\ast(T^{0,1}X)$  and $\Sigma^i\in\Omega^2(\Sigma)\otimes\Phi^\ast(T^{1,0}X)$~.
 \end{itemize} 
  The $B$ model possesses two supercharges. The first supercharge acts on the fields as follows:
\begin{equation}
  \begin{aligned}\label{transfb1}
      \delta\phi^i&=0,\qquad\delta\rho^i=-2\ii\dd\phi^i,\qquad \delta\Sigma^{i}=2\dd^\Gamma\rho^i+\frac{\ii}{2}R^i{}_{j\bar{l}k}\eta^{\bar l}\rho^j\wedge\rho^k~,\\
      \delta\phi^{\bar i}&=\eta^{\bar i},\qquad\delta\eta^{\bar i}=0,\qquad \delta\theta^{\bar i}=\beta^{\bar i}-\Gamma^{\bar i}_{\bar{j}\bar{k}}\eta^{\bar j}\theta^{\bar k},\qquad \delta\beta^{\bar i}=-\Gamma^{\bar i}_{\bar{j}\bar{k}}\eta^{\bar j}\beta^{\bar k}~,
  \end{aligned}
\end{equation}
where $\dd^\Gamma \rho^i=\dd\rho^i+\Gamma^i_{j k}\dd\phi^j \wedge\rho^k$.
In order to match with the standard presentation of the $B$ model we can use instead $\theta_i=g_{i \bar j} \theta^{\bar j}$ and $\beta_i=g_{i \bar j} \beta^{\bar j}$ transforming as:
\begin{align}
    \delta \theta_i= \beta_i~,\qquad \delta \beta_i= 0~.
    \end{align}
This supercharge is nilpotent, $\delta^2=0$. The action on the fields of the second supercharge $\hat \delta$  is presented in appendix~\ref{app:susy}. This second supercharge is also nilpotent and anticommutes with the first hence the complete superalgebra reads
\begin{equation}
    \delta^2=0~,\quad \hat \delta^2=0~, \quad\{\delta,\hat \delta\}=0~.
\end{equation}

In the absence of  a superpotential the Lagrangian reads
\begin{equation}\label{eq--L.Bmodel}
  \mathcal{L}_\mathrm{D}=2g_{i\bar j}\dd\phi^i\wedge\star\dd\phi^{\bar j}-\frac{1}{2}\Sigma^i\beta_i-\ii g_{i\bar j}\rho^i\wedge\star\dd^\Gamma\eta^{\bar j}-\dd^\Gamma\rho^i\theta_i-\frac{\ii}{4}R^i{}_{j\bar l k}\rho^j\wedge\rho^k\eta^{\bar l}\theta_i~.
\end{equation}
This Lagrangian is $\delta$-exact. Indeed, $\mathcal{L}_\mathrm{D}=\delta   \mathcal{V}$ where
\begin{equation}\label{eq--VandVh}
  \mathcal{V}=\ii g_{i\bar j}\rho^i\wedge\star\dd\phi^{\bar j}-\frac{1}{2}\Sigma^i\theta_i~.
\end{equation}
${\cal L}_{D}$ is also $\hat \delta$-exact as shown in~\eqref{lagdeltahat}.
Let  the superpotential $W$ be a holomorphic function of the $\phi^i$ and $\widetilde W$ be its complex conjugate. The corresponding Lagrangians read
\begin{align}
& {\cal L}_{W}= \frac{1}{2}\partial_m W \Sigma^m+ \frac{i}{4} D_j \partial_k W \rho^j\wedge \rho^k~,\label{eq--W}\\
& {\cal L}_{\widetilde W}=\frac{1}{2}  \delta \hat \delta \left(\widetilde W \right)~,
    \end{align}
where we define $D_j \partial_k W=\partial_j\partial_k W-\Gamma^i_{j k} \partial_i W $. We remark that $ {\cal L}_{\widetilde W}$ is $\delta$-exact (and $\hat\delta$-exact). In contrast ${\cal L}_{W}$, while supersymmetric, is not a $\delta$-variation.

The $B$ model is topological. To see this we can change the metric on $\Sigma_g$ keeping constant the cohomological variables. The dependence of $\mathcal{L}_\mathrm{D}$ and ${\cal L}_{\widetilde W}$ on the metric on $\Sigma_g$  is $\delta$-exact. As for ${\cal L}_{W}$ it does not depend explicitly on the metric on $\Sigma_g$.
We remark the following:
\begin{itemize}
\item The auxiliary fields can be integrated out leading to
$$
\Sigma^i= g^{i\bar j}\partial_{\bar j} \widetilde W\star\! 1~,\qquad \beta^{\bar i}=g^{j\bar i}\partial_j W~.
$$
 After elimination of the auxiliary fields $\mathcal{L}_\mathrm{D}$ is no longer $\delta$-exact. As a result this somewhat obscures the topological nature of the model. The action of $\delta$ after elimination of the auxiliary fields is:
\begin{equation}
  \begin{aligned}\label{transfb2}
      \delta\phi^i&=0~,\qquad\delta\rho^i=-2\ii\dd\phi^i~,\\
      \delta\phi^{\bar i}&=\eta^{\bar i}~,\qquad\delta\eta^{\bar i}=0~,\qquad \delta\theta_i=\partial_i W~,
  \end{aligned}
\end{equation} 
 which still closes off-shell.

\item The auxiliary two-form $\Sigma^i$ is often dualized into a scalar (see e.g.~\cite{Labastida:1991qq}). This also obscures the topological nature of the model because if $\Sigma^i$ is kept constant under changes of the metric the same is not true for the dual scalar. We will see that, in the context of the equivariant $B$ model, it is natural to have the auxiliary field $\Sigma^i$ be a two-form on $\Sigma_g$.
\end{itemize}

\subsection{Observables and localization}

Supersymmetric local operators in the $B$ model are given by holomorphic functions $f(\phi)$. These are $\delta$-closed by virtue of $\delta\phi^i=0$. Correlators of these operators,
\begin{equation}
 \label{corrl}
 \langle \prod_{\alpha} f_\alpha(\phi (x_\alpha) ) \rangle~,
\end{equation}
 are independent of their positions $x_\alpha$ on the worldsheet because $\dd\phi^i=\frac{\ii}{2}\delta\rho^i$ is $\delta$-exact. Consider now functions of the form $u^i (\phi)\partial_iW(\phi)$ with  $u$ being 
 a holomorphic vector field $u\in T^{1,0}X$. According to \eqref{transfb2}
\begin{equation}
f(\phi)u^i\partial_iW(\phi)=f(\phi)u^i\delta\theta_i=\delta(f(\phi)u^i\theta_i)~,
\end{equation}
 so that functions containing $\partial_iW$ as a factor are $\delta$-exact. Consequently, operators in the $\delta$-cohomology are given by the ring $\mathcal{R}$ of holomorphic functions on $X$ modulo the ideal generated by $\partial_1W,\dots,\partial_NW$ ($\dim_{\mathbb{C}}X=N$). On flat space $X=\mathbb{C}^N$, this ring is simply the polynomial ring
\begin{equation}
    \mathcal{R}=\mathbb{C}[\phi^1,\dots,\phi^N]/(\partial_1W,\dots,\partial_NW),
\end{equation}
known as the chiral ring of the theory for a generic $W$.

Our interest is in computing the correlators~\eqref{corrl} of generators $f_\alpha$ of the chiral ring. For this purpose we can use localization. We consider the following Lagrangian:
\begin{equation}\label{eq--L.loc}
  \mathcal{L}= \mathcal{L}_\mathrm{D}+\mathcal{L}_W+\ell \tilde{\mathcal{L}}_{\tilde{W}}~,
\end{equation}
where $\ell$ is a large parameter that multiplies $\delta$-exact terms.
The bosonic terms in this Lagrangian are given by
\begin{equation}
\begin{aligned}
    \mathcal{L}_\mathrm{bos}=&~\left(2g_{i\bar j}\dd\phi^i\wedge\star\dd\phi^{\bar j}-\frac{1}{2}\Sigma^i\beta_i\right) +{1\over 2}(\partial_i W)\Sigma^i+{\ell\over 2} (\partial_{\bar i}\widetilde{W})\beta^{\bar i}\star 1~.
\end{aligned}
\end{equation}
The auxiliary fields can be integrated out:
\begin{equation}
\label{auxint}
    \Sigma^i=\ell g^{i \bar j}\left(\partial_{\bar j}\widetilde{W}\right)\star 1,\qquad \beta^{\bar i}=g^{j \bar i}\left(\partial_{ j}{W}\right).
\end{equation}
This results in the bosonic Lagrangian\footnote{This is the same result obtained in~\cite{Vafa:1990mu} by rescaling the worldsheet metric~.}
\begin{equation}
    \begin{aligned}
        \mathcal{L}_\mathrm{bos}=&~2 g_{i\bar j}\dd\phi^i\wedge\star\dd\phi^{\bar j}+\frac{\ell}{2}g^{i\bar j}\left(\partial_{ i}{W}\right)\wedge\star\left(\partial_{\bar j}\widetilde{W}\right).
     \end{aligned}
\end{equation}
In the large-$\ell$ limit the path integral localizes to configurations where the $\phi^i$ are constant and equal to a critical point of the superpotential:
\begin{equation}
 \phi^i=\phi^i_\mu~,\quad (\partial_i W)(\phi_\mu)=0~,   
\end{equation}
with the Greek index $\mu$ labeling critical points.

Provided the critical points are isolated and non-degenerate, which is true generically, we expand the scalar fields around each critical point so that $\phi^i=\phi^i_\mu+{1\over \sqrt{\ell}} \Delta\phi^i~.$ The path integral is then one-loop exact. The fermionic and bosonic contributions to the one-loop determinant cancel up to the contributions of   constant modes. The fields $\rho^i$ have $g$ zero-modes where $g$ is the genus of $\Sigma_g$. The fields $\theta^{\bar i}, \eta^{\bar i}$ have one zero-mode each as do $\Delta \phi^i,~\Delta \phi^{\bar i}$. Hence the zero-modes give the one-loop determinant
\begin{equation}\label{one-loop-Bm-gen}
{(\det \partial_{\bar i}\partial_{\bar j} \widetilde W(\bar \phi_\mu))  (\det \partial_i\partial_j W(\phi_\mu))^{g}\over (\det \partial_{\bar i}\partial_{\bar j} \widetilde W(\bar \phi_\mu))(\det \partial_i\partial_j W(\phi_\mu))}=  (\det \partial_i\partial_j W(\phi_\mu))^{g-1}~,
\end{equation}
and the correlators~\eqref{corrl} evaluate to~\cite{Vafa:1990mu}
\begin{equation}
\label{Bmodcorr}
\sum_\mu  (\det \partial_i\partial_j W(\phi_\mu))^{g-1} \prod_{\alpha} f_\alpha(\phi_\mu)~.
\end{equation}
These are independent of $\widetilde W$ as expected because it only enters $\delta$-exact terms in the Lagrangian. They are also explicitly independent of the insertion positions. From now on we will consider the case $g=0$, $\Sigma_0=S^2$.

The correlators~\eqref{Bmodcorr} have a nontrivial dependence on the one-loop determinants. However the structure of the chiral ring is insensitive to them. Let the $f_\alpha(\phi)$ form a basis for the chiral ring of the theory.  Generically, when the critical points are isolated and non-degenerate, the $f_\alpha(\phi_{\mu}) \equiv f_\alpha^\mu$ are encoded in an invertible square matrix $f_\alpha^\mu$. The topological metric and three-point functions can be written as
\begin{equation}
\eta_{\alpha\beta}=\langle f_\alpha(\phi) f_\beta(\phi)\rangle=\sum_\mu N_\mu f^\mu_\alpha f^\mu_\beta~,\qquad
C_{\alpha\beta\gamma}=\langle f_\alpha(\phi) f_\beta(\phi)f_\gamma(\phi)\rangle=\sum_\mu N_\mu f^\mu_\alpha f^\mu_\beta f^\mu_\gamma~,
\end{equation}
where $N_\mu$ is the one-loop determinant (\ref{one-loop-Bm-gen}) with $g=0$. The structure constants for the chiral ring are then
\begin{equation}
\label{stconfr}
C^{\alpha}_{\beta\gamma}=\eta^{\alpha \kappa}C_{\kappa \beta \gamma}=\sum_\mu f_\beta^\mu f_\gamma^\mu (f^{-1})^\alpha_\mu~,
\end{equation}
and are independent of the one-loop contributions $N_\mu$. Hence deformations of the superpotential that do not change the critical points can modify the topological metric and various correlators but not the structure constants. As an example consider a superpotential $W$ which is an analytic function of a single $\mathbb{C}$ valued field $\phi$. 
The one-loop determinant gives then 
\begin{equation}
    N_\mu^{-1}= W''(\phi_\mu)~.
\end{equation}

Define a new superpotential $\widehat W$ as follows:
\begin{equation}
\widehat{W}(\phi) = \int\limits_0^\phi \partial_z W (z) \e{- F(z)}~dz~.
\end{equation}
Then 
\begin{equation}
    \partial \widehat{W} =  (\partial W) \e{- F}~,
\end{equation}  
so that the critical points of $W$ and $\widehat W$ coincide.
The one-loop determinant now gives 
\begin{equation}
    N_\mu^{-1}= \widehat{W}''(\phi_\mu) =  W''(\phi_\mu) \e{- F(\phi_\mu)}~.
\end{equation}  
Hence correlators computed with the $\widehat{W}$ superpotential differ from the ones computed with $W$. We could have obtained the same result by adding to the action an additional supersymmetric term of the form $\int F(\phi)$~.
While the correlators are different the structure of the chiral ring is the same. The structure constants computed via~\eqref{stconfr} coincide in both cases as they are independent of the $N_\mu$~. 

\section{The equivariant \texorpdfstring{$B$}{B} model}
\label{sec:equivB}

We now consider an equivariant version of the $B$ model whose supersymmetry transformations were introduced in \cite{Yagi:2014toa} and \cite{Nekrasov:2018pqq}. 
The relation between a ${\cal N}=(2,2)$ theory in rigid curved superspace~\cite{Closset:2014pda} and this model was considered in~\cite{Closset:2015rna} and is the subject of Appendix~\ref{app:NLSM}. 
We focus on the theory on $S^2$, where it is equivariant with respect to the $U(1)$-action given by azimuthal rotations.
More precisely, introducing polar coordinates $\vartheta\in [0,\pi]$ and $\varphi\sim \varphi+2\pi$, the metric on $S^2$ has the form
\bea
\label{metsq}
ds^2= R^2 f(\vartheta)^2( d\vartheta^2 +\sin(\vartheta)^2 d\varphi^2)~.
\eea 
Here $f(\vartheta)$ is a smooth strictly positive function. The round sphere of radius $R$ is obtained with the choice $f(\vartheta)=1$. Introducing the real equivariant parameter  $\epsilon$, the azimuthal isometry corresponds to translations in $\varphi$ and is generated by $v={\epsilon\over R} {\partial\over \partial\varphi}$. We also make use of the one-form $\kappa=\epsilon R f(\vartheta)^2 \sin(\vartheta)^2d\varphi $ related to $v$ by contraction with the metric.

\subsection{Supersymmetry and Lagrangians}
\label{susylageq}
Equivariance deforms the supercharges discussed in the previous section  so that their algebra reads:
\begin{align}
\delta^2=-2\ii {\cal L}_v~, \quad\hat\delta^2=2\ii {\cal L}_v~,\quad \{\delta, \hat \delta\}=0~,
\end{align}
where ${\cal L}_v$ is the action of the Lie derivative along $v$ on the fields of the model.

The action of the supersymmetry variation $\delta$ on the fields is deformed to
\begin{equation}
  \begin{aligned}\label{transfb}
      \delta\phi^i&=\iota_v\rho^i,\qquad\delta\rho^i=-2\ii\dd\phi^i-\ii\iota_v\Sigma^i-\Gamma^i_{jk}\iota_v\rho^j\rho^k,\\
\delta\Sigma^{i}&=2\dd^\Gamma\rho^i+\Gamma^{i}_{jk}\iota_v\Sigma^j\wedge\rho^k+\frac{\ii}{2}R^i{}_{j\bar{l}k}\eta^{\bar l}\rho^j\wedge\rho^k~,\\
      \delta\phi^{\bar i}&=\eta^{\bar i},\qquad\delta\eta^{\bar i}=-2\ii\iota_v\dd\phi^{\bar i},\qquad
      \delta\theta^{\bar i}=\beta^{\bar i}+2\star(\kappa\wedge\dd\phi^{\bar i})-\Gamma^{\bar i}_{\bar{j}\bar{k}}\eta^{\bar j}\theta^{\bar k},\\
      \delta\beta^{\bar i}&=-2\ii\iota_v\dd^\Gamma\theta^{\bar i}-2\star(\kappa\wedge\dd^\Gamma\eta^{\bar i})-\Gamma^{\bar i}_{\bar{j}\bar{k}}\eta^{\bar j}\beta^{\bar k}+R^{\bar i}_{\bar{j}l\bar{k}}\iota_v\rho^l\eta^{\bar j}\theta^{\bar k},
  \end{aligned}
\end{equation}
where $\dd^\Gamma \rho^i=\dd\rho^i+\Gamma^i_{j k}\dd\phi^j \wedge\rho^k$,~ $\dd^\Gamma \eta^{\bar i}=\dd\eta^{\bar i}+\Gamma^{\bar i}_{\bar j\bar k}\dd\phi^{\bar j} \eta^{\bar k}$ and $\dd^\Gamma \theta^{\bar i}=\dd \theta^{\bar i}+\Gamma^{\bar i}_{\bar j\bar k}\dd\phi^{\bar j} \theta^{\bar k}$~.
We also quote the transformations of $\theta_i = g_{i\bar j} \theta^{\bar j}$ and $\beta_i= g_{i\bar j} \beta^{\bar j}$:
\begin{align}
    \delta \theta_i&= \beta_i +2 g_{i \bar j}\star(\kappa\wedge\dd\phi^{\bar j}) +\Gamma^k_{i j} \iota_v\rho^{j} \theta_k~,\\ 
    \delta \beta_i&= -2 \ii \iota_v\dd^\Gamma \theta_i-2 \star(\kappa\wedge\dd^\Gamma\eta_i)+\Gamma^{k}_{i j} \iota_v\rho^j \beta_{ k}+R_{i \bar j  j \bar k} \iota_v\rho^j \eta^{\bar j} \theta^{\bar k}~.    
\end{align}
One important difference with respect to the non-equivariant case is that the auxiliary field $\Sigma^i$ now enters the variation of $\rho^i$. 

Inspecting~\eqref{transfb} it would seem appropriate to simplify the action of $\delta$ by redefining the auxiliary field $\beta^{\bar i}$  as follows
\begin{equation}
    \tilde \beta^{\bar i}=\beta^{\bar i}+2 \star (\kappa \wedge \dd \phi^{\bar i})~.
\end{equation}
We prefer to use $\beta^{\bar i}$ instead of $\tilde \beta^{\bar i}$ for two reasons. Firstly the relation to the physical ${\cal N}=(2,2)$ theory as described in appendix~\ref{twichi} is more direct in terms of $\beta^{\bar i}$. In second instance, when we will discuss Lagrangians, expressions in terms of $\beta^{\bar i}$ are simpler and the relation between the equivariant and non-equivariant models is more transparent. The action of the second supersymmetry variation $\hat\delta$ in the equivariant setting is presented in appendix~\ref{app:susy}.

We can summarize the transformations~\eqref{transfb} and make equivariance explicit by working with superfields. Let $\rho^i=\rho^i_a\xi^a$ and $\Sigma^i=\Sigma^i_{a b}\xi^a\xi^b$ with $\xi^a$ Grassmann variables and $\dd=\xi^a\partial_a$, $~\kappa=\kappa_a \xi^a$. We can then define the superfields
\begin{align}
\label{sfields}
&{\bf \Phi}^i=\phi^i+\rho^i+{\ii}\left(\Sigma^i+{i\over 2} \Gamma^{i}_{j k}\rho^j\rho^k\right)~,\\ 
& {\bf H}^{\bar i}= \eta^{\bar i}+2\ii d\phi^{\bar i}~, \\  
& {\bf B}^{\bar i}=\beta^{\bar i}+2\star(\kappa\wedge d\phi^{\bar i})-\Gamma^{\bar i}_{\bar j \bar k}\eta^{\bar j} \theta^{\bar k}+2\ii d \theta^{\bar i}~.\label{sfieldsf}
\end{align}
In order to describe the action of $\delta$ on these superfields we define the equivariant differential ${\cal D}_v=-2\ii\xi^a\partial_a-v^a{\partial\over \partial \xi^a}$ which satisfies:
\begin{equation}
    {\cal D}_v ^2=-2\ii {\cal L}_v~.
\end{equation}
In terms of ${\cal D}_v$ the superfield variations then read,
\begin{equation}
\label{eq:sf.transf}
  \delta{\bf \Phi}^i={\cal D}_v{\bf \Phi}^i~,\quad \delta {\bf H}^{\bar i}={\cal D}_v {\bf H}^{\bar i}~,\quad   \delta {\bf B}^{\bar i}={\cal D}_v {\bf B}^{\bar i}~.
\end{equation}
This shows that using a two-form auxiliary field $\Sigma^i$ is natural in the context of the equivariant $B$ model. 

In order to obtain a $\delta$-invariant Lagrangian consider again ${\mathcal V}$ as in~\eqref{eq--VandVh},
\begin{equation}\label{eq--VandVhn}
  \mathcal{V}=\ii g_{i\bar j}\rho^i\wedge\star\dd\phi^{\bar j}-\frac{1}{2}\Sigma^i\theta_i~.
\end{equation}
Its $\delta$-variation yields $\delta\mathcal{V}=\mathcal{L}_\mathrm{D}$ with
\begin{equation}\label{eq--L.Bmodeleq}
  \mathcal{L}_\mathrm{D}=2g_{i\bar j}\dd\phi^i\wedge\star\dd\phi^{\bar j}-\frac{1}{2}\Sigma^i\beta_i-\ii g_{i\bar j}\rho^i\wedge\star\dd^\Gamma\eta^{\bar j}-\dd^\Gamma\rho^i\theta_i-\frac{\ii}{4}R^i{}_{j\bar l k}\rho^j\wedge\rho^k\eta^{\bar l}\theta_i~,
\end{equation}
being precisely the standard $B$ model Lagrangian \eqref{eq--L.Bmodel} without equivariance. In appendix \ref{twichi} we show that this can easily be understood in terms of twisting a ${\cal N}=(2,2)$ theory.
This indifference to equivariant deformations implies that the standard $B$ model is invariant under the family of supersymmetries parametrized by the free parameter $\epsilon $. Note that ${\cal L}_\mathrm{D}$ is also $\hat\delta$-exact as shown in~\eqref{lagdeltahat}.

One important difference with respect to the non-equivariant case arises when integrating out the auxiliary fields. Consider the action~\eqref{eq--L.Bmodeleq}. The auxiliary fields can be integrated out by setting $\Sigma^i=0,~ \beta^{\bar i}=0$.
Substituting back into the $\delta$-variations~\eqref{transfb} these do not close off shell anymore unlike in the non-equivariant case. This will be the case also after deforming the action with the observables we describe in the next section.

\subsection{Observables}
\label{sec:observables}

Here we will consider the supersymmetric observables of the equivariant $B$ model we introduced in section~\ref{susylageq}. There are two qualitative differences with respect to the non-equivariant case. Firstly, supersymmetry restricts the observables to be inserted on $v$-invariant cycles. For instance, local observables ${\cal O}_0$ are supersymmetric only when inserted at the poles of the sphere. Secondly, equivariance allows supersymmetric observables to depend non-trivially on the latitude $\vartheta$. For instance, the superpotential $W$ can be promoted to a holomorphic function of the $\phi^i$ which also depends explicitly on $\vartheta$. 

Let $\omega_0\in\Omega^0(S^2)$ and $\omega_2\in\Omega^2(S^2)$ such that $\iota_v \omega_2=2\ii \dd \omega_0$. 
This implies $\iota_v d \omega_0=0$. From these forms we can build $ \Omega=\omega_0-\omega_{2 a b} \xi^a\xi^b$ which is equivariantly closed (i.e. ${\cal D}_v\Omega=0$). A very general class of observables can be defined in terms of the superfields ${\bf \Phi}^i$ and ${\bf H}^{\bar i}$ in~\eqref{sfields} and $\Omega$. For a $(0,k)$ form on $X$ we can define
\begin{equation}
   \Pi_A= A({\bf \Phi},\bar \phi)_{{\bar i}_1,{\bar i}_2,\ldots {\bar i}_k} {\bf H}^{{\bar i}_1} {\bf H}^{{\bar i}_2}\ldots  {\bf H}^{{\bar i}_k}~,
\end{equation}

Now let $A$ be a $\bar \partial$-closed $(0,k)$-form on $X$.  We can then consider $\Omega\,\Pi_A$.
The action of $\delta$ on $\Omega\,\Pi_A$ is given by the equivariant differential ${\cal D}_v$. By expanding $\Omega\,\Pi_A$ in components as ${\cal O}_0+{\cal O}_{1 a}\xi^a-{\cal O}_{2 a b}\xi^a\xi^b$ we obtain observables satisfying the descent equations
\begin{equation}
 \begin{aligned}\label{eq:descent}
\delta\mathcal{O}_0&=\iota_v\mathcal{O}_1,\qquad \delta\mathcal{O}_1=\iota_v\mathcal{O}_2-2\ii\dd\mathcal{O}_0,\qquad \delta\mathcal{O}_2=-2\ii\dd\mathcal{O}_1~.
  \end{aligned}
\end{equation}
Hence the insertion in the path integral of $ \mathcal{O}_0\vert_p$ where $p$ is one of the two poles $\vartheta=0$ or $\vartheta=\pi$ is invariant under supersymmetry. Similarly $\int_{S^2} \mathcal{O}_2$ is invariant under supersymmetry and can be used as part of a supersymmetric action. The choice $\omega_0=1$ with $\omega_2=0$ reproduces the superpotential term~\eqref{eq--W} with  $W=F$.
We can also obtain a $\delta$-invariant insertion by integrating $ \mathcal{O}_1$ over a $v$-invariant cycle $C$ on $S^2$.  Because we are considering a worldsheet with trivial fundamental group this insertion is $\delta$-exact. Indeed we can consider integrating ${\cal O}_2$ on a hemisphere $D$ bounded by $C$ and take a $\delta$-variation to obtain:
\begin{equation}
    \delta \int_D {\cal O}_2=-2\ii\int_D d{\cal O}_1=-2\ii \int_C {\cal O}_1~.
\end{equation}
If the $(0,k)$-form $A$ is $\bar\partial$-exact, $A=\bar\partial B$, we have 
\begin{equation}\Omega\,\Pi_A=\delta ( \Omega\, \Pi_B ) -{\cal D}_v(  \Omega\, \Pi_B)~.
\end{equation}
Similarly if $\Omega= {\cal D}_v (\Lambda)={\cal D}_v (\omega_{1 a}\xi^a)$ where $\omega_1=\omega$ is $v$-invariant (that is $\iota_v d\omega_1+d\iota_v \omega_1=0$). We have 
\begin{equation}
\Omega\,\Pi_A=\delta (\Lambda \Pi_A)-{\cal D}_v (\Lambda \Pi_A)~.
\end{equation}
Therefore, in these two cases, integrating the corresponding ${\cal O}_0\,~{\cal O}_1,~{\cal O}_2$ over $v$-invariant cycles gives rise to  $\delta$-exact quantities.  It follows that these observables are classified by $(A, \Omega)\in H^{0,\bullet}(X) \times H^\bullet_{equiv} (S^2)$. 

The simplest case, and the one we will consider in detail in the following, is obtained taking $A=F(\bf\Phi)$ to be a holomorphic function of ${\bf \Phi}$. Then the corresponding superfield expression reduces to ${i\over 2}\Omega\,F({\bf \Phi})$ on which $\delta$ acts as the equivariant differential ${\cal D}_v$. By expanding in components we define the three observables
\begin{equation}\label{eq--O}
  \begin{aligned}
    \mathcal{O}_0&=\ii{\omega_0\over 2} F(\phi)~,\\
    \mathcal{O}_1&=\ii {\omega_0\over 2} \partial_i F(\phi)\rho^i~,\\
    \mathcal{O}_2&={\omega_0\over 2} \partial_i F(\phi)\Sigma^i+\frac{\ii}{4}\omega_0D_i\partial_j F(\phi)\rho^i\wedge\rho^j+\ii{\omega_2 \over 2}F(\phi)~,
  \end{aligned}
\end{equation}
which satisfy the descent equations \eqref{eq:descent}.

\paragraph{More general dependence on latitude.}
\label{genob}
In the equivariant setting we can introduce observables whose dependence on the latitude $\vartheta$ is more general. Indeed consider a function $W(\phi,x)$ where $x$ is the position on the worldsheet. Let's introduce the notation $\partial W \equiv \partial_x W dx $ where we only take the derivative with respect to the explicit\footnote{This is to be compared with $dW=\partial W+\partial_i W \partial\phi^i$~.}  $x$-dependence of $W$. Then we can define the observables:
\begin{equation}
\begin{aligned}
\label{genobsdf}
   & {\cal O}_0= {\ii\ifthenelse{\boolean{eps}}{\epsilon}{}\over 2} W(\phi,x)~,\\
   & {\cal O}_1={\ii\ifthenelse{\boolean{eps}}{\epsilon}{}\over 2} \partial_i W(\phi,x) \rho^i,\\
   &  {\cal O}_2= {\ifthenelse{\boolean{eps}}{\epsilon}{1}\over 2}\partial_i W(\phi,x) \Sigma^i+{\ifthenelse{\boolean{eps}}{\ii\epsilon}{\ii}\over 4} \, D_j\partial_k W(\phi,x)\rho^j\wedge\rho^k-{\ifthenelse{\boolean{eps}}{\epsilon}{}\kappa\over ||v||^2}\wedge \partial W(\phi,x)~.
 \end{aligned}
 \end{equation}
Provided that $\partial W$ vanishes sufficiently fast at the poles the last term in ${\cal O}_2$ is well-defined.
We require that $\iota_v \partial W=0$, hence $W$ is explicitly only a function of the latitude $\vartheta$. It then follows that~\eqref{genobsdf} satisfy the descent equations as before:
\begin{align}
    \delta {\cal O}_0=\iota_v {\cal O}_1~,\quad \delta {\cal O}_1=\iota_v {\cal O}_2-2\ii \dd {\cal O}_0~,\quad \delta {\cal O}_2=-2\ii \dd {\cal O}_1~.
\end{align}
The previous observables are a special case of these  where the $\vartheta$-dependence factorizes.

At the fixed points of the action of $v$, that is at the two poles $\vartheta=0$ or $\vartheta=\pi$, the $\delta$-variation of ${\cal O}_0$ in~\eqref{eq--O} or~\eqref{genobsdf} vanishes. Hence the insertion of ${\cal O}_0$ at the poles is supersymmetric. This is to be contrasted with the non-equivariant case where chiral operators $F(\phi)$ are supersymmetric when inserted anywhere on $S^2$ and correlators of these insertions are independent of position. As we reviewed, in the non-equivariant case with superpotential $W$, chiral operators $F(\phi)$ are equivalent modulo $\partial_i W$, thus giving rise to a finitely generated chiral ring. In the equivariant case the notion of a chiral ring is more subtle as the superpotential $W(\phi,\vartheta)$ can now depend explicitly on $\vartheta$. 
In the next section we will calculate correlators of holomorphic functions $F(\phi)$ inserted at the poles of the sphere. As we will see insertions at the $\vartheta=0$ pole are equivalent modulo $\partial_i W(\phi,0)$ while insertions at the $\vartheta=\pi$ pole are equivalent modulo $\partial_i W(\phi,\pi)$. Hence the chiral rings at the two poles can in principle differ. We also give a formal derivation of this result in appendix \ref{app:chiral}. 

\paragraph{On-shell observables.}

There is another class of observables that we can consider. They are of the form
\begin{equation}
\label{eq:O.cs}
  \mathcal{O}_0=A^{i_1\dots i_p}{}_{\bar j_1\dots\bar j_q}(\phi,\bar\phi,\vartheta)\eta^{\bar j_1}\dots\eta^{\bar j_q}\theta_{i_1}\dots\theta_{i_p}~,
\end{equation}
where $A$ is $\bar\partial$-closed.
We can again find $\mathcal{O}_1$, $\mathcal{O}_2$ which are related to $\mathcal{O}_0$ through descent relations. However, in contrast to the observables introduced previously, here the descent relations hold only on-shell (see \cite{Labastida:1994ss} for a discussion in the non-equivariant setting):
\begin{align*}
  \delta\mathcal{O}_0=\iota_v\mathcal{O}_1+(\mathrm{e.o.m.}),\quad\delta\mathcal{O}_1=\iota_v\mathcal{O}_2-2\ii\dd\mathcal{O}_0+(\mathrm{e.o.m}),\quad\delta\mathcal{O}_2=-2\ii\dd\mathcal{O}_1+(\mathrm{e.o.m})~,
\end{align*}
where $(\mathrm{e.o.m})$ denotes terms that are proportional to the equations of motion. In particular, following \cite{Witten:1991zz,Labastida:1994ss}, write these terms in $\delta\mathcal{O}_2$ as $\frac{\delta S}{\delta\Psi_K}\cdot\zeta_K$, where $\Psi_K$ denotes the fields in the theory and the $\zeta_K$ are some functions of these. We can then consider deforming the action as
\begin{equation}
  S+t\int_{S^2}\mathcal{O}_2~,
\end{equation}  
where $t$ is some small parameter.
The deformed action is not $\delta$ invariant. However up to order $t^2$ it is invariant under the modified supersymmetry transformation
\begin{equation}
\label{eq:delta.cs}
  \check\delta\Psi_K=\delta\Psi_K+t\zeta_K~.
\end{equation}
Computing $\mathcal{O}_2$ and $\zeta_K$ for $\mathcal{O}_0$ of the general form \eqref{eq:O.cs} is laborious and we do not present its form here. Instead, we will consider an example with $p=q=1$ in section \ref{sec:deformations}. For this case the deformation can be made to work at all orders in $t$.

\section{Superpotentials}
\label{sec:superpotentials}

In this section we consider the equivariant $B$ model with $X=\mathbb{C}^N$ and a superpotential $W(\phi,\vartheta)$ depending explicitly on the latitude $\vartheta$ and compute correlators of supersymmetric local operators using localization. We find that the localization locus is not given by constant maps. Instead, the locus comprises maps that connect critical points of $W(\phi,0)$ to critical points of $W(\phi,\pi)$. 
By inspecting the localization result for the correlators, we see that local supersymmetric observables at the $\vartheta=0$ pole satisfy the chiral ring relations deriving from $W(\phi,0)$ while those at the $\vartheta=\pi$ pole satisfy the chiral ring relations deriving from $W(\phi,\pi)$. 

\subsection{Localization}

Here we consider localizing the equivariant $B$ model with a superpotential \begin{equation}
    W(\phi,\vartheta)=W_1(\phi)+\cos(\vartheta)W_2(\phi)~.
\end{equation} Superpotentials with more general dependence on the latitude are considered in appendix~\ref{Apploc}.
Our starting point is the $\delta$-exact Lagrangian~\eqref{eq--L.Bmodeleq} to which we add two observables. Consider the formula for ${\cal O}_2$ in~\eqref{eq--O}. For the first observable we choose $\omega_0=\ifthenelse{\boolean{eps}}{\epsilon}{1}$ and $F=W_1$, for the second one $\omega_0=\ifthenelse{\boolean{eps}}{\epsilon}{}\cos(\vartheta)$ and $F=W_2$. The corresponding bosonic terms are 
\begin{equation}
\begin{aligned}
    \mathcal{L}_\mathrm{bos}=&~\left(2g_{i\bar j}\dd\phi^i\wedge\star\dd\phi^{\bar j}-\frac{1}{2}\Sigma^i\beta_i\right)\\
    &+{\ifthenelse{\boolean{eps}}{\epsilon}{1}\over 2}(\partial_i W_1 +\cos(\vartheta)\partial_i W_2)\Sigma^i  +{\ii\over 2} W_2\, \omega_2~,
\end{aligned}
\end{equation}
where $\omega_2=2i R \ifthenelse{\boolean{eps}}{}{\epsilon^{-1}}\sin(\vartheta)\dd\vartheta \wedge \dd\varphi~.$

In order to proceed with localization we add several $\delta$-exact terms proportional to a large real positive parameter $\ell$
\begin{equation}
{\cal L}_{\rm loc}={\ell\over 2}\delta\left( ||v||^2{\cal V}+\ifthenelse{\boolean{eps}}{\epsilon}{}\partial_{\bar i}(\widetilde W_1+\cos(\vartheta)\widetilde W_2)\star\theta^{\bar i} +i\ifthenelse{\boolean{eps}}{\epsilon}{} \partial_{i}( W_1+\cos(\vartheta) W_2)\kappa\wedge\rho^i \right)~.
\end{equation}
Here ${\cal V}$ is defined in~\eqref{eq--VandVhn}. Notice that while these localization terms are $\delta$-exact they are not $\hat\delta$-invariant. 

With these additions the bosonic Lagrangian is 
\begin{equation}\label{eq:Lfull}
\begin{aligned}
    \mathcal{L}_\mathrm{bos}=&~ (1+\ell||v||^2)\left(2g_{i\bar j}\dd\phi^i\wedge\star\dd\phi^{\bar j}-\frac{1}{2}\Sigma^i\beta_i\right)\\
    &+{\ifthenelse{\boolean{eps}}{\epsilon}{1}\over 2}(1+\ell ||v||^2)(\partial_i W_1 +\cos(\vartheta)\partial_i W_2)\Sigma^i+{\ifthenelse{\boolean{eps}}{\epsilon}{1}\over 2} \ell(\partial_{\bar i}\widetilde{W}_1+\cos(\vartheta)\partial_{\bar i}\widetilde{W}_2)\beta^{\bar i}\star 1\\
    & +{i\over 2}W_2\,\omega_2+\ell\ifthenelse{\boolean{eps}}{\epsilon}{} \kappa\wedge (\dd{\widetilde W}_1+\cos(\vartheta)\dd {\widetilde W}_2)+\ell\ifthenelse{\boolean{eps}}{\epsilon}{} \kappa\wedge( \dd W_1+\cos(\vartheta)  \dd W_2)~.
\end{aligned}
\end{equation}
The auxiliary fields can be integrated out:
\begin{equation}
\label{auxinta}
    \Sigma^i=\frac{\ell}{1+\ell||v||^2}g^{i \bar j}\left(\partial_{\bar j}\widetilde{W}_1+\cos(\vartheta) \partial_{\bar j}\widetilde{W}_2\right)\star 1,\qquad \beta^{\bar i}= g^{j \bar i}\left(\partial_{ j}{W}_1+\cos(\vartheta) \partial_{ j}{W}_2\right)~,
\end{equation}
which results in the bosonic Lagrangian
\begin{align}
        \mathcal{L}_\mathrm{bos}=&~2 (1+\ell ||v||^2)g_{i\bar j}\dd\phi^i\wedge\star\dd\phi^{\bar j}+\frac{\ell\ifthenelse{\boolean{eps}}{\epsilon}{}}{2 }g^{i\bar j}(\partial_{ i}{W}_1+\cos(\vartheta) \partial_{ i}{W}_2)\wedge\star(\partial_{\bar j}\widetilde{W}_1+\cos(\vartheta) \partial_{\bar j}\widetilde{W}_2)\nonumber\\
          & +{i\over 2}W_2\,\omega_2+\ell\ifthenelse{\boolean{eps}}{\epsilon}{} \kappa\wedge  (\dd{\widetilde W}_1+\cos(\vartheta) \dd{\widetilde W}_2)+\ell\ifthenelse{\boolean{eps}}{\epsilon}{} \kappa\wedge (\dd W_1+\cos(\vartheta) \dd W_2).
     \end{align}
Next we recast all the terms proportional to the large parameter $\ell$ into a sum of squares
 \begin{align}
 \label{sumofsquares}
        &\mathcal{L}_\mathrm{bos}=~2  g_{i\bar j}\dd\phi^i\wedge\star\dd\phi^{\bar j}  +{i\over 2}W_2\,\omega_2 +2\ell  g_{i\bar j}\iota_v \dd \phi^i \iota_v\dd \phi^{\bar j} \star 1\\
        &+2\ell   g_{i \bar j}\left(\star(\kappa \wedge \dd\phi^{\bar j})+{\ifthenelse{\boolean{eps}}{\epsilon}{1} \over 2 }\partial^{\bar j}\left({W}_1+\cos(\vartheta){W}_2\right)\right)\left(\kappa\wedge \dd\phi^{i}+{\ifthenelse{\boolean{eps}}{\epsilon}{1}\over 2 }\partial^{i}\! \left(\widetilde{W}_1+\cos(\vartheta)\widetilde{W}_2\right)\star 1\right). \nonumber    
\end{align}
From this expression we see that the functional integral localizes to scalar field configurations that are invariant along $v$ so that $\iota_v \dd \phi^{\bar i}=0$ and satisfy the differential equation
\begin{equation}
\label{eqtrka}
 \star (\kappa\wedge \dd \phi^{\bar i})+{\ifthenelse{\boolean{eps}}{\epsilon}{1}\over 2 }\,g^{j\bar i} \left(\partial_{ j}{W}_1+\cos(\vartheta) \partial_{ j}{W}_2\right)=0~.
\end{equation}
These two conditions are the BPS conditions $\delta\eta^{\bar i}=0$ and $\delta \theta^{\bar i}=0$
\begin{equation}
\iota_v \dd \phi^{\bar i}=0~,\qquad \beta^{\bar i}+2\star(\kappa\wedge\dd\phi^{\bar i})=0~,
\end{equation}
with $\beta^{\bar i}$ given by~\eqref{auxinta}. The remaining BPS condition, namely $\delta \rho^i=0$ reads
 \begin{equation}
 \delta\rho^i\propto2\dd\phi^i+\iota_v\Sigma^i={2\over 1+\ell ||v||^2}  \dd \phi^i~,
 \end{equation}
where we used~\eqref{auxinta} and~\eqref{eqtrka}. As we will discuss below, for regular solutions of~\eqref{eqtrka}  $\dd \phi^i$ vanishes at the poles.  Hence in the large-$\ell$ limit $\delta\rho^i$ vanishes as well.

Similarly we can localize in the presence of the more general observables described in \eqref{genobsdf}. The localization Lagrangian is considered in Appendix~\ref{Apploc}. The localization locus is then given by field configurations that satisfy  
\begin{equation}
\label{eqtrtwo}
\iota_v d\phi^{\bar i}=0~,\qquad \star (\kappa\wedge \dd \phi^{\bar i})+{1\over 2 }\,g^{j\bar i} \left(\partial_{ j}{W(\phi, \vartheta)}\right)=0~.
\end{equation}
In this case the superpotential $W(\phi,\vartheta)$ includes an explicit dependence on $\vartheta$.  By taking $W= \ifthenelse{\boolean{eps}}{\epsilon(W_1+\cos(\vartheta) W_2)}{W_1+\cos(\vartheta) W_2}$ we reproduce~\eqref{eqtrka}.

Finally, evaluating the action corresponding to~\eqref{sumofsquares} on the localization locus we obtain (see Appendix~\ref{Apploc} for details)
 \begin{equation}
  S=-\ifthenelse{\boolean{eps}}{}{{1\over \epsilon}}\int_{S^2} d\varphi \wedge dW=-\ifthenelse{\boolean{eps}}{2\pi}{{2\pi\over \epsilon}} \int_0^\pi {d\over d\vartheta}W d\vartheta= \ifthenelse{\boolean{eps}}{2\pi}{{2\pi\over \epsilon}} (W|_{\vartheta=0}-W|_{\vartheta=\pi})~.
\end{equation}

\paragraph{Solutions to the BPS equations}

Consider now the form of the solutions to~\eqref{eqtrka}. To start we analyze their  behavior near the poles.
By using that $\kappa=\epsilon R f(\vartheta)^2\sin^2(\vartheta) d\varphi$ we rewrite~\eqref{eqtrka}  as
\begin{equation}
\label{eqtrb}
 {\ifthenelse{\boolean{eps}}{}{\epsilon}\over R} \sin(\vartheta) \partial_\vartheta \phi^{\bar i}-{1\over 2}\,g^{j\bar i} \left(\partial_{ j}{W}_1+\cos(\vartheta) \partial_{ j}{W}_2\right)=0~.
\end{equation}
The behavior of regular solutions at the poles is 
\begin{equation}
\label{condpole}
\begin{aligned}
&{\rm for~} \vartheta\sim 0: ~~~ \phi^i=\phi^i_N + O(\vartheta^2)~~{\rm where~~}\partial_i(W_1+W_2)(\phi^i_N)=0~,\\
&{\rm for~} \vartheta\sim \pi: ~~~ \phi^i=\phi^i_S + O((\pi-\vartheta)^2)~~{\rm where~~}\partial_i(W_1-W_2)(\phi^i_S)=0~.
\end{aligned}
\end{equation}
Hence these solutions interpolate between the critical points of $W_1+W_2$ at $\vartheta=0$ and the critical points of $W_1-W_2$ at $\vartheta=\pi$~.
Solutions that diverge at the poles do not have a finite action and can be neglected in the following. 
The regular solutions are BPS because $\dd \phi^i$ vanishes at the poles so that $\delta \rho^i=0$ everywhere in the large $\ell$ limit.

Determining the exact solutions to~\eqref{eqtrb} is beyond our ability. However, there are some interesting limiting cases that we list below.

In the special case in which a subset of the critical points of $W_1$ and $W_2$ coincide there are constant solutions corresponding to these critical points. A specific instance is $W_2=0$. The equation then reduces to a particular case of those studied in~\cite{Cecotti:1992rm} that considered domain wall solutions of
\begin{equation}
 \partial_x \phi^{\bar i}-{\alpha} {g^{j\bar i}\over 2} \partial_{ j}{W}=0~,\qquad x\in\mathbb{R}
\end{equation}
where $\alpha$ is a choice of phase. After the substitution $x={\ifthenelse{\boolean{eps}}{ }{\epsilon }\over 2}\log(\tan(\vartheta/2))$ our equation corresponds to $\alpha=1$. If we fix the phase $\alpha$ to this value the analysis of~\cite{Cecotti:1992rm} tells us that for generic choices of $W_1$ the only solutions are constants at each of the critical points of $W_1$. There can be domain wall solutions interpolating between two critical points $\phi^i_N$ and $\phi^i_S$ if the difference $W_1(\phi_N)-W_1(\phi_S)$ is real which makes $W_1$ non-generic.

Another interesting limit is that of $R \rightarrow \infty$ in which case the solutions track the critical points of $W_1+\cos(\vartheta) W_2$ as $\vartheta$ varies from one pole to the other~\footnote{There can also be exceptional domain wall solutions as in the case $W_2=0$ but we don't expect them to contribute to the correlators.}. For large $R$ these solutions should get slightly modified and we will refer to them as tracking solutions.  In the limit $R \rightarrow\nolinebreak  \infty$ the more general equation~\eqref{eqtrtwo} has tracking solutions as well. They are approximately given by $\phi^i(\vartheta)$ satisfying 
\begin{equation}
(\partial_i W)(\phi(\vartheta),\vartheta)=0~.
\end{equation}
We can distinguish different situations:
\begin{itemize}
\item{The critical points of $W(\phi,\vartheta)$ at each value of $\vartheta$ are non-degenerate, distinct and their number is independent of $\vartheta$. Then for large $R$ there are tracking solutions that map each critical point at one pole to a critical point at the other pole. We will refer to this as the tame case.}
\item{If the number of critical points of $W(\phi,\vartheta)$ changes with $\vartheta$ then either some of the critical points merge or go to infinity. In the latter case the corresponding tracking solution diverges and does not give rise to a saddle for the localization computation. The case where some of the solutions merge is not generic and can arise at the boundary of distinct tame cases. We will comment on this further in section~\ref{commentsex}~.} 
\end{itemize}

\subsection{Correlators}
Unlike for the standard $B$ model insertions of holomorphic functions of the $\phi^i$ anywhere on $S^2$ are not supersymmetric. However they are supersymmetric when inserted at the $\vartheta=0$ or $\vartheta=\pi$ poles on the sphere.  Hence we will consider the following correlators:
\begin{equation}
\label{corrdef}
\langle \prod_\alpha f_\alpha(\phi (0))\prod_\beta f_\beta(\phi (\pi)) \rangle~.
\end{equation}
We now turn to study these correlators in different cases. We will see that the insertions at the $\vartheta=0$ and $\vartheta=\pi$ poles have the structure dictated by the chiral ring of $W(\phi,\vartheta=0)$ and $W(\phi,\vartheta=\pi)$ respectively. 

\subsubsection{Standard superpotential: \texorpdfstring{$W_2=0$}{W2=0}}
For $W_2=0$ the Lagrangian is identical to that in the $B$ model up to $\delta$-exact terms. Hence the result for the correlators is the same as well. The localization locus includes constant maps sitting at the critical points $\phi^i_\mu$ of the superpotential $W_1$. These  give a contribution to the correlators of the form
\begin{equation}
\sum_\mu  (\det \partial_i\partial_j W(\phi_\mu))^{-1} \prod_{\alpha} f_\alpha(\phi_\mu)~,
\end{equation}
which coincides with~\eqref{Bmodcorr} for $g=0$~.

As we have discussed, for special choices of $W_1$ there can be one or more domain wall solutions interpolating between different critical points. These are not generically present. From the general discussion above these saddles should not contribute to the correlators. Indeed we expect the exceptional saddles to have both bosonic and fermionic zero-modes. Because of the fermionic zero-modes  the contribution of these configurations to the correlators then vanishes. One  example is considered in Appendix \ref{Appdom}.

\subsubsection{Position dependent superpotential: tracking solutions}
\label{subsub:tracking}

For large radius $R$ and for generic choices of $W(\phi,\vartheta)$ the only solutions should be the tracking solutions described above. Let there be multiple such solutions $\phi^i=\phi^i_\mu$ labeled by $\mu$. We will denote the values that these solutions attain at the poles as
\begin{equation}
\phi^i_{\mathrm{N}, \mu}=\phi^i_\mu(\vartheta=0)~,\quad \phi^i_{\mathrm{S}, \mu}=\phi^i_\mu(\vartheta=\pi)~.
\end{equation}
These are critical points for $W_N=W(\phi,\vartheta=0)$ and  $W_S=W(\phi,\vartheta=\pi)$ respectively.
The localization result for the correlators~\eqref{corrdef} is then given by:
\begin{equation}
\label{restrck}
\sum_\mu N_\mu \left(\e{\ifthenelse{\boolean{eps}}{2\pi}{{2\pi\over \epsilon}}W(0,\phi_{\mathrm{N}, \mu})}\prod_\alpha f_\alpha(\phi_{\mathrm{N}, \mu}) \right)\left(\e{-\ifthenelse{\boolean{eps}}{2\pi}{{2\pi\over \epsilon}}W(\pi,\phi_{\mathrm{S}, \mu})} \prod_\beta f_\beta(\phi_{\mathrm{S}, \mu} ) \right)~,
\end{equation}
where $N_\mu$ is the one-loop contribution around the tracking solutions. $N_\mu$ should not depend on $\widetilde{W}$ and the expectation coming from equivariance is that it factorizes into two contributions from the two poles. 
Here we do not determine the one-loop contributions $N_\mu$, which we leave to future work. Nevertheless we will see that, as in the non-equivariant case there are some quantities that do not depend on them.

\paragraph{The tame case.}

We will first consider the case where each critical point of $W(\phi,0)$ is connected to a critical point of $W(\phi,\pi)$ by a tracking solution.

For two operators inserted at the same pole, say N, the correlator gives 
\begin{equation}
\label{eq:top.metric}
\nu_{\mathrm{N}, \alpha\beta} =\langle f_\alpha(\phi (0)) f_\beta(\phi(0))\rangle = \sum_\mu N_\mu \e{\ifthenelse{\boolean{eps}}{2\pi}{{2\pi\over \epsilon}}W(\phi_{\mathrm{N}, \mu},0)}  \e{-\ifthenelse{\boolean{eps}}{2\pi}{{2\pi\over \epsilon}}W(\phi_{\mathrm{S}, \mu},\pi)} f_\alpha(\phi_{\mathrm{N}, \mu})f_\beta(\phi_{\mathrm{N}, \mu})~.
\end{equation}
This is of the same form as the topological metric for the superpotential $W(\phi,0)$ in the non-equivariant case. Indeed we can write it as $\sum_\mu {\tilde N}_\mu  f_\alpha(\phi_{\mathrm{N}, \mu})f_\alpha(\phi_{\mathrm{N}, \mu})$ with non-vanishing ${\tilde N}_\mu$ (in the absence of zero-modes).
For three operators inserted at the north pole the correlator gives
\begin{equation}
\label{eq:C}
\mathcal{C}_{\mathrm{N},\alpha\beta\gamma}=\sum_\mu N_\mu\e{\ifthenelse{\boolean{eps}}{2\pi}{{2\pi\over \epsilon}}W(\phi_{\mathrm{N}, \mu},0)}  \e{-\ifthenelse{\boolean{eps}}{2\pi}{{2\pi\over \epsilon}}W(\phi_{\mathrm{S}, \mu},\pi)}f_\alpha(\phi_{\mathrm{N},\mu})f_\beta(\phi_{\mathrm{N},\mu})f_\gamma(\phi_{\mathrm{N},\mu}).
\end{equation}
Again this is of the same form than the three-point function in the non-equivariant case.
Similarly for more insertions at the same pole the result will be of the same form as for the corresponding correlator in the non-equivariant model.
Hence the $N$ pole correlators have the structure dictated by the chiral ring of $W(\phi,0)$. Similarly for the $S$ pole correlators and the chiral ring of $W(\phi,\pi)$. We can then choose bases of generators for the two rings which we will denote $f_\alpha^N(\phi)$ and $f_\alpha^S(\phi)$. When the critical points $\phi_{N,\mu}$ and $\phi_{S,\mu}$ are non-degenerate and isolated the matrices $(f^N)^\mu_\alpha=f_\alpha^N(\phi_{N,\mu})$ and $(f^S)^\mu_\alpha=f_\mu^S(\phi_{S,\mu})$ are invertible.

With this in mind we can turn to the two-point function where the two insertions are on opposite poles:
\begin{equation}
\label{eq:map.NL}
\psi_{\alpha\beta} =\langle f^N_\alpha(\phi (0)) f^S_\beta(\phi (\pi))\rangle = \sum_\mu N_\mu \e{\ifthenelse{\boolean{eps}}{2\pi}{{2\pi\over \epsilon}}W(\phi_{\mathrm{N}, \mu},0)}  \e{-\ifthenelse{\boolean{eps}}{2\pi}{{2\pi\over \epsilon}}W(\phi_{\mathrm{S}, \mu},\pi)}  f^N_\alpha(\phi_{\mathrm{N}, \mu})f^S_\beta(\phi_{\mathrm{S}, \mu})~.
\end{equation}
We can then define ${\psi^\alpha}_\beta=(\nu^{-1}_\mathrm{N})^{\alpha \kappa}\psi_{\kappa \beta}= \sum_\mu  (f^{-1}_\mathrm{N})^\alpha_\mu (f_\mathrm{S})_\alpha^\mu $ which is independent of the one-loop contributions.
This matrix and its inverse can be used to relate $N$ and $S$ pole correlators, for instance:
\begin{equation}
\label{eq:map.g}
  (\nu_\mathrm{N})_{\alpha\beta} {\psi^\alpha}_\kappa{\psi^\beta}_\rho=(\nu_\mathrm{S})_{\kappa\rho}~.
\end{equation}
Because it only depends on the $(f_\mathrm{S})_\alpha^\mu$ and $(f_\mathrm{N})_\alpha^\mu$ the matrix $\psi$ also provides a map between the structure constants for the operators inserted at the two poles:
\begin{equation}
\label{eq:map.C}
  {(\psi^{-1})^\kappa}_\alpha ({C_\mathrm{N}})^\alpha_{\beta \gamma} {\psi^\beta}_\rho{\psi^\gamma}_\lambda=(C_\mathrm{S})^\kappa_{\rho \lambda}~.
\end{equation}
Let us denote by $\mathcal{R}_{N,S}$ the chiral  rings corresponding to $W(\phi,\vartheta=0)$ and $W(\phi,\vartheta=\pi)$. In particular, \eqref{eq:map.C} implies that the map 
\begin{equation}
\label{eq:psi}
\psi:\mathcal{R}_N\rightarrow\mathcal{R}_S,\;f^N_\alpha\mapsto (\psi^{-1})^\beta{}_\alpha f^S_\beta~,
\end{equation}
(after extending by linearity) is an algebra isomorphism.

\paragraph{Large deformation.} Consider now the case when the number of critical points of $W(\phi,0)$ is strictly less than that of $W(\phi,\pi)$. Then the regular tracking solutions originating at $\vartheta=0$ will connect to a subset of the critical points at $\vartheta=\pi$. 
As a consequence the $(f_\mathrm{N})^\mu_\alpha$ are a square invertible matrix but not the $(f_\mathrm{S})^\mu_\alpha$. We can still define \begin{equation}
    {\psi^\alpha}_\beta=(\nu^{-1}_\mathrm{N})^{\alpha \kappa}\mu_{\kappa \beta}= \sum_\mu  (f^{-1}_\mathrm{N})^\alpha_\mu (f_\mathrm{S})_\beta^\mu~,
\end{equation} but not its inverse.
What happens is that, because only a subset of the critical points of $W(\phi,\pi)$ is connected by a tracking solution, the ``effective" superpotential at $\vartheta=\pi$ is not $W(\phi,\pi)$. The chiral ring of the ``effective" superpotential at $\vartheta=\pi$ is strictly smaller than that of $W(\phi,\pi)$ and isomorphic to that of $W(\phi,0)$.

\subsubsection{Comments on \texorpdfstring{$Q$}{Q}-exactness}
\label{commentsex}
Consider the expression~\eqref{restrck} for the correlators~\eqref{corrdef}. Apart from the one-loop contribution it depends on the values $\phi^i_{\mathrm{N}, \mu}$ and $\phi^i_{\mathrm{S}, \mu}$ that the tracking solutions attain at the two poles. These are critical points for $W_\mathrm{N}=W(\phi,\vartheta=0)$ and  $W_\mathrm{S}=W(\phi,\vartheta=\pi)$. Let's consider the case in which the tracking solutions provide a map between the critical points of $W_\mathrm{N}$ and $W_\mathrm{S}$ which is bijective. There are multiple such maps differing by permutations of the $\phi^i_{\mathrm{S}, \mu}$ associated to the $\phi^i_{\mathrm{N}, \mu}$. The correlators~\eqref{corrdef} depend on the permutation as does the matrix ${\psi^\alpha}_\beta$. Suppose we have two $W(\phi,\vartheta)$ with the same $W_\mathrm{N}$ and $W_\mathrm{S}$ whose tracking solutions give rise to a map between critical points that differ by a permutation of the $\phi^i_{\mathrm{S}, \mu}$. Deforming one of the two $W(\phi,\vartheta)$ into the other should be a $\delta$-exact deformation as $W_\mathrm{N}$ and $W_\mathrm{S}$ are unchanged. However the correlators~\eqref{corrdef} and the matrix ${\psi^\alpha}_\beta$ change under the deformation. We see that this happens because the structure of the BPS locus is not continuous under such a deformation. At the boundary between different patterns for the tracking solutions some of them merge together. An example displaying this behavior is considered below.

\subsubsection{Examples}

\paragraph{Tame case.} As a first example we consider a minimal model from the $A$-series with one-dimensional flat target $X=\mathbb{C}$. Specifically, consider the position-dependent superpotential
\begin{equation}
  W(\phi,\vartheta)={1\over 4}\phi^4-e^{3 \ii \alpha{ \vartheta\over \pi}}\phi~,
\end{equation}
where $\alpha\in\mathbb{R}$ is a deformation parameter.
As discussed earlier, equivariance requires local observables to sit at the two poles of the sphere. Consequently, we obtain a chiral ring at each pole:
\begin{equation}
\mathcal{R}_\mathrm{N}=\mathbb{C}[\phi]/(\phi^3-1)~,\qquad\mathcal{R}_\mathrm{S}=\mathbb{C}[\phi]/(\phi^3-\e{3\ii\alpha})~.
\end{equation}
The superpotential at the two poles is a deformation of the quartic minimal model. The critical points of $W$ at the north, resp. south pole are given by
\begin{equation}
  \{\phi_{\mathrm{N},\mu}\}_\mu=\{1,\e{2\ii\pi/3},\e{4\ii\pi/3}\}~,\qquad\{\phi_{\mathrm{S,\mu}}\}_\mu=\{\e{\ii\alpha},\e{2\ii\pi/3+\ii\alpha},\e{4\ii\pi/3+\ii\alpha}\}~.
\end{equation}
The critical points at the two poles are connected in the following way by the tracking solutions \eqref{eqtrb}:
\begin{equation}
  \label{eq:example.tracking}
  1\rightsquigarrow\e{\ii\alpha}~,\qquad \e{2\pi\ii/3}\rightsquigarrow\e{2\pi\ii/3+\ii\alpha}~,\qquad \e{4\pi\ii/3}\rightsquigarrow\e{4\pi\ii/3+\ii\alpha}~.
\end{equation}
For both rings we can choose the following generators:
\begin{equation}
\label{eq:basis.mm}
  f_1\equiv1~,\qquad f_2(\phi)=\phi~,\qquad f_3(\phi)=\phi^2~.
\end{equation}
With this choice of basis we easily compute  \eqref{eq:top.metric} at both poles:
\begin{align}
  (\nu_{\mathrm{N},\alpha\beta})=&\begin{pmatrix}
    \tilde N_1+\tilde N_2+\tilde N_3 & M_1 & M_2\\ 
    M_1 & M_2 & \tilde N_1+\tilde N_2+\tilde N_3\\
    M_2 & \tilde N_1+\tilde N_2+\tilde N_3 & M_1
  \end{pmatrix},\\[.5em]
  (\nu_{\mathrm{S},\alpha\beta})=&\begin{pmatrix}
    \tilde N_1+\tilde N_2+\tilde N_3 & M^{(\alpha)}_1 & M^{(\alpha)}_2\\
    M^{(\alpha_1)} & M^{(\alpha)}_2 & \e{3\ii\alpha}(\tilde N_1+\tilde N_2+\tilde N_3)\\
    M^{(\alpha)}_2 & \e{3\ii\alpha}(\tilde N_1+\tilde N_2+\tilde N_3) & M^{(\alpha)}_3
  \end{pmatrix},
\end{align}
where we have defined
\begin{equation}
  M^{(\alpha)}_1=\e{\ii\alpha}(\tilde N_1+\e{2\pi\ii/3}\tilde N_2+\e{4\pi\ii/3}\tilde N_3),\qquad M^{(\alpha)}_2=\e{2\ii\alpha}(\tilde N_1+\e{4\pi\ii/3}\tilde N_2+\e{2\pi\ii/3}\tilde N_3)
\end{equation}
and $M^{(\alpha)}_3=\e{3\ii\alpha}M^{(\alpha)}_1$ with $M_1=M^{(0)}_1$, $M_2=M^{(0)}_2$. The one-loop factors are given by $\tilde N_\mu=N_\mu\e{{2\pi\over \epsilon}W(0,\phi_{\mathrm{N}, \mu})}  \e{-{2\pi\over \epsilon}W(\pi,\phi_{\mathrm{S}, \mu})}$ with $N_\mu$ the one-loop determinant evaluated on the tracking solutions. Similarly, we can use \eqref{eq:C} to compute $\mathcal{C}_{\mathrm{N},\alpha\beta\gamma}$, $\mathcal{C}_{\mathrm{S},\alpha\beta\gamma}$. It is then a simple exercise to verify the relations
\begin{equation}
  \nu_{\mathrm{N},\alpha\kappa}(C_\mathrm{N})^\kappa{}_{\beta\gamma}=\mathcal{C}_{\mathrm{N},\alpha\beta,\gamma}~,\qquad \nu_{\mathrm{S},\alpha\kappa}(C_\mathrm{S})^\kappa{}_{\beta\gamma}=\mathcal{C}_{\mathrm{S},\alpha\beta\gamma}~,
\end{equation}
where $(C_{\mathrm{N,S}})^\alpha{}_{\beta\gamma}$ are the structure constants of $\mathcal{R}_{\mathrm{N,S}}$.

We can now consider the two-point functions $\psi_{\alpha\beta}$ introduced in \eqref{eq:map.NL} which map between $\mathcal{R}_\mathrm{N}$ and $\mathcal{R}_\mathrm{S}$. In fact, we proceed directly to compute $\psi^\alpha{}_\beta$ which is independent of the one-loop factors:
\begin{equation}
  (\psi^\alpha{}_\beta)=\begin{pmatrix}
    1 & 0 & 0 \\
    0 & \e{\ii\alpha} & 0 \\
    0 & 0 & \e{2\ii\alpha}
  \end{pmatrix}.
\end{equation} 
It is now an easy exercise to check the identities \eqref{eq:map.g} and \eqref{eq:map.C} explicitly.

Note that for $\alpha=2\pi k/3$ ($k\in\mathbb{Z}$) we have $\mathcal{R}_\mathrm{N}=\mathcal{R}_\mathrm{S}$. However, for $k\notin 3\mathbb{Z}$ the tracking solutions connect the critical points non-trivially (cf. \eqref{eq:example.tracking}). We see that for these choices of $k$ the two-point functions $\nu_{\alpha\beta}$ and $\psi^\alpha{}_\beta$ are \textit{not} equivalent to the case of $k\in3\mathbb{Z}$. Instead, the map $\psi$ introduced in \eqref{eq:psi} provides two inequivalent algebra automorphisms for $k=1\mmod3$ and $k=2\mmod 3$. 

\paragraph{Large deformation.}
As a second example we consider a superpotential $W(\phi,\vartheta)$ such that the critical points are obtained by solving
\begin{equation}
0=W'(\phi,\vartheta)=(1+\phi^2)(\sin^2(\vartheta/2)\phi+\alpha(\phi-1)-1)~,
\end{equation}
where $\alpha$ is a parameter.
Two of the critical points $\phi=\pm \ii$ are independent of $\vartheta$. The third critical point is at $\phi=1$ for $\vartheta=\pi$ and $\phi=1+\alpha^{-1}$ for $\vartheta=0$. For $\alpha$ not too small and large $R$ there is a tracking solution connecting these values. As $\alpha\rightarrow 0$ the tracking solution would diverge at $\vartheta=0$. We see that this is related to the fact that for $\alpha=0$ the superpotential is cubic at $\vartheta=0$ and hence only has the $\phi=\pm \ii$ critical points. Correlators with powers of $\phi$ inserted at $\vartheta=0$ are generically singular for $\alpha\rightarrow 0$. 
Denoting the algebra generators by
\begin{equation}
\label{eq:basis.mma }
  f_1\equiv1~,\qquad f_2(\phi)=\phi~,\qquad f_3(\phi)=\phi^2~,
\end{equation}
we can define 
\begin{equation}
{\psi^\alpha}_\beta=(\nu^{-1}_\mathrm{N})^{\alpha\kappa}\psi_{\kappa \beta}= \sum_\mu  (f^{-1}_\mathrm{N})^\alpha_\mu (f_\mathrm{S})_\beta^\mu
\end{equation}
which is independent of one-loop determinants.
We get
\begin{equation}
  (\psi^\alpha{}_\beta)=\begin{pmatrix}
    1 & 0 & -1 \\
    0 & 1 & 0 \\
    0 & 0 & 0
  \end{pmatrix}-{\alpha\over 1+2\alpha+2\alpha^2}\begin{pmatrix}
    0 & 1 & -2\alpha \\
    0 & 0 & 0 \\
    0 & 1 & -2\alpha
  \end{pmatrix}.
\end{equation} 
Hence ${\psi^\alpha}_\beta$ is finite in the limit $\alpha\rightarrow 0$. However at $\alpha=0$ the matrix ${\psi^\alpha}_\beta$ is not invertible. 

\section{Complex structure deformations}
\label{sec:deformations}

In this section we consider the equivariant $B$ model with deformations to the complex structure of the target $X$. Such deformations in the standard $B$ model setting have been considered in \cite{Labastida:1994ss}. Here, because of equivariance, we can allow for more general deformations depending explicitly on the latitude $\vartheta$ of $S^2$. This can be interpreted as a theory on a family of Calabi-Yau manifolds $X$ whose complex structure is parametrized by $\vartheta$. We present this setting in some detail below.

\subsection{Families of complex structures}

While $X$ is always Calabi-Yau elsewhere in this work, for this subsection $X$ can be any complex manifold. It is equipped with an integrable almost complex structure $J\in\End(TX)$ associated with the decomposition $TX\otimes\mathbb{C}=T^{1,0}X\oplus T^{0,1}X$ of the complexified tangent space. Now consider a different almost complex structure $\widetilde J$ which we think of as a small deformation of $J$ in the following sense: $\pi^{0,1}$ -- the projection to the second summand -- provides an isomorphism between $T^{0,1}X$ and $\widetilde{T}^{0,1}X$. Consequently, for fixed $J$, $\widetilde J$ gives rise to a linear map
\begin{equation}
  \mu:T^{0,1}X\overset{(\pi^{0,1})^{-1}}{\longrightarrow}\widetilde{T}^{0,1}X\overset{\pi^{1,0}}{\longrightarrow}T^{1,0}X,
\end{equation}
i.e. $\mu\in\Hom(T^{0,1}X,T^{1,0}X)\simeq\Omega^{0,1}_X(T^{1,0}X)$. Thus, small deformations in $J$ are characterised by elements in $\Omega^{0,1}_X(T^{1,0}X)$ and vice versa.

More generally, we can consider a family of deformations of the complex structure on $X$. These are then represented by elements $\mu(t)\in\Omega^{0,1}_{X}(T^{1,0})$ with $t=(t^1,\dots,t^m)\in U$ a set of complex parameters and $U\subset \mathbb{C}^m$ a polydisc around the origin, such that $\mu(0)=0$. In order for $\mu(t)$ to give rise to an integrable complex structure it has to satisfy the Kodaira-Spencer (KS) equation:
\begin{equation}
\label{eq:KS.formal}
  \bar\partial_X\mu(t)+\frac{1}{2}[\mu(t),\mu(t)]=0~.
\end{equation}
Then $\{\mu(t)\}_{t\in U}$ gives a family of complex structures on $X$ varying smoothly\footnote{In fact, one can obtain a holomorphically varying family by imposing the additional requirement that $\mu(t)$ depend on $t$ holomorphically. Then $\mu\in\pr^\ast_X\Omega^{0,1}_{X}(T^{1,0}X)$ gives rise to a complex structure on $X\times U$. However, in our setting we only care about pointwise integrability in $U$.} with $t$. In local holomorphic coordinates $\{z^i\}$ on $X$ we can write $\mu(t)=\frac{\ii}{2}\mu^i{}_{\bar j}(z,t)\dd z^{\bar j}\otimes\frac{\partial}{\partial z^i}$. Then the KS equation becomes
\begin{equation}
  \label{eq:KS}
    \partial_{[\bar i}\mu^k{}_{\bar j]}-\frac{\ii}{2}\partial_j\mu^k{}_{[\bar i}\mu^j{}_{\bar j]}=0~.
\end{equation}
At a point $(z,t)\in X\times U$ the new tangent space $\widetilde T_z^{0,1}X$ is spanned by
\begin{equation}
  \frac{\partial}{\partial z^{\bar i}}+\frac{\ii}{2}\mu^j{}_{\bar i}(z,t)\frac{\partial}{\partial z^j}~,
\end{equation} 
where $\bar i,j=1,\dots,\dim_{\mathbb{C}} X$.

So far we have discussed finite deformations $\mu(t)$ of $J$. One can instead consider infinitesimal deformations and ask when these can be integrated to $\mu(t)$. For this purpose, let $w\in T^{1,0}_{0} U$. We define the infinitesimal deformation $w\cdot\mu\in\Omega^{0,1}_X(T^{1,0}X)$ by differentiating $\mu(t)$ along $w$ and evaluating at $t=0$. Then 
\begin{equation}
  \bar \partial(w\cdot\mu)=w\cdot\bar\partial_X\mu=-[w\cdot\mu,\mu(0)]=0~,
\end{equation}
and we see that $w\cdot\mu$ is an element of $H^{0,1}(X;T^{1,0}X)$. Now we can formally write $\mu(t)$ as a power series in $t$,
\begin{equation}
\label{eq:series}
\mu(t)=\mu^{(1)}+\mu^{(2)}+\mu^{(3)}+\mu^{(4)}+\dots\;.
\end{equation}
with coefficients $\mu^{(i)}$ in $\Omega^{0,1}_X(T^{1,0}X)$ that are homogeneous in $t$ of degree $i$. If $\{\beta_\alpha\}_{\alpha=1,\dots,m}$ form a basis of $H^{0,1}(X;T^{1,0}X)$, we can write $\mu^{(1)}=\beta_\alpha t^\alpha$.
Now plugging this form of $\mu(t)$ in \eqref{eq:KS.formal} gives a recursive relation for the $\mu^{(i)}$. In general, there will be obstructions to a solution of these relations. However, in the case of $X$ being Calabi-Yau it was proved in \cite{Tian:1987,Todorov:1989umc} that a unique such solution for $\mu(t)$ of the form \eqref{eq:series} exists.
Hence, in the special case of $X$ being Calabi-Yau small deformations of the complex structure are in one-to-one correspondence with elements in $H^{0,1}(X;T^{1,0}X)$.

In this work we want to consider deformations of the complex structure that depend on the position on the worldsheet (specifically, on $\vartheta$). This can be achieved by considering the trivial fibration $Y=X\times S^2 \overset{\pi}{\longrightarrow} S^2$ with $X$ as fibre and $S^2$ as the base. Then $Y$ is an almost complex manifold (since $X$ and $S^2$ are) and we can equip it with an almost complex structure $\Xi\in\End(TY)$. In particular, we choose $\Xi$ such that is does not follow from the product structure of $Y$. By imposing that for every $x\in S^2$
\begin{equation}
  \Xi|_{T\pi^{-1}(x)}\in\End(T\pi^{-1}(x))~,
\end{equation}
$\Xi$ restricts on each fibre $X_x:=\pi^{-1}(x)$ (which is diffeomorphic to $X$) to a complex structure $J(x):=\Xi|_{T\pi^{-1}(x)}$. In this way we obtain a family $\{X_x\}_{x\in S^2}$ of complex manifolds, all diffeomorphic to $X$. By choosing $\Xi$ to be close enough to the product almost complex structure, $J(x)$ constitute small deformations of the original complex structure as discussed earlier. These can be described precisely by $\mu(t)$ introduced above, where now the complex deformation parameters $t$ are to be understood as smooth functions of $\vartheta$, $t=t(\vartheta)$.

\subsection{Deformed supersymmetry and localization}

Let us now look at the observable in \eqref{eq:O.cs} for $p=q=1$, $\mathcal{O}_0=A^{i}{}_{\bar j}(\phi,\bar\phi,\vartheta)\eta^{\bar i}\theta_j$. For fixed $\vartheta$ this corresponds precisely to an element in $H^{0,1}(X;T^{1,0}X)$ and thus to an infinitesimal deformation of the complex structure on $X$. Then adding $\mathcal{O}_2$ to the equivariant $B$ model Lagrangian $\mathcal{L}_\mathrm{D}$ simply corresponds to accounting for the change in $\mathcal{L}_\mathrm{D}$ due to deformations by $A=\mu^{(1)}$ (the first term in \eqref{eq:series}). As described in section \ref{sec:observables}, the resulting action is no longer invariant under the original supersymmetry variations $\delta$, but instead is invariant to first order in $t$ under some modified variations \eqref{eq:delta.cs} involving $\mu^{(1)}$. 

We would like to describe finite deformations $\mu$ in our theory, invariant under some modified supersymmetry transformations $\tilde\delta$ to all orders in $t$. For the non-equivariant case it was pointed out in \cite{Labastida:1994ss} that this can be achieved simply by replacing $\mu^{(1)}$ with $\mu$ in the supersymmetry transformations, satisfying the KS equation \eqref{eq:KS.formal}. Then the new Lagrangian accounting for finite deformations is obtained as $\mathcal{L}=\tilde\delta\mathcal{V}$. We proceed in complete analogy and obtain the deformed supersymmetry algebra
\begin{equation}
  \begin{aligned}\label{eq--SUSY.deformed}
  \tilde\delta\phi^i=&~\iota_v\rho^i+\frac{\ii}{2}\mu^i{}_{\bar j}\eta^{\bar j}~,\qquad\tilde\delta\rho^i=-2\ii\dd\phi^i-\ii\iota_v\Sigma^i-\Gamma^i_{jk}\iota_v\rho^j\rho^k-\mu^i{}_{\bar j}\dd\phi^{\bar j}-\frac{\ii}{2}\partial_k\mu^i{}_{\bar j}\rho^k\eta^{\bar j}~,\\  
\tilde\delta\Sigma^{i}=&~2\dd^\Gamma\rho^i+\Gamma^{i}_{jk}\iota_v\Sigma^j\wedge\rho^k+\frac{\ii}{2}R^i{}_{j\bar{l}k}\eta^{\bar l}\rho^j\wedge\rho^k+\frac{\ii}{2}\partial_k\mu^i{}_{\bar j}\eta^{\bar j}\Sigma^k+\ii D_k\mu^i{}_{\bar j}\rho^k\wedge\dd\phi^{\bar j}\\
  &-\frac{1}{4}D_lD_k\mu^i{}_{\bar j}\eta^{\bar j}\rho^l\wedge\rho^k-\ii {\kappa\over ||v||^2}\wedge \partial \mu^i_{\bar j}\eta^{\bar j}~,\\
  \tilde\delta\phi^{\bar i}=&~\eta^{\bar i}~,\qquad\tilde\delta\eta^{\bar i}=-2\ii\iota_v\dd\phi^{\bar i}~,\qquad \tilde\delta\theta^{\bar i}=\beta^{\bar i}+2\star(\kappa\wedge\dd\phi^{\bar i})-\Gamma^{\bar i}_{\bar{j}\bar{k}}\eta^{\bar j}\theta^{\bar k}~,\\ 
  \tilde\delta\beta^{\bar i}=&-2\ii\iota_v\dd^\Gamma\theta^{\bar i}-2\star(\kappa\wedge\dd^\Gamma\eta^{\bar i})-\Gamma^{\bar i}_{\bar{j}\bar{k}}\eta^{\bar j}\beta^{\bar k}+R^{\bar i}_{\bar{j}l\bar{k}}(\iota_v\rho^l\eta^{\bar j}\theta^{\bar k}+\frac{\ii}{2}\mu^l{}_{\bar r}\eta^{\bar r}\eta^{\bar j}\theta^{\bar k})~,
  \end{aligned}
\end{equation}
where $\partial\mu$ denotes the exterior derivative\footnote{Note that this object transforms tensorially under diffeomorphisms of $X$ (see discussion in previous subsection). This would no longer be the case for non-trivial fibrations $Y\rightarrow\Sigma$, where the $\phi^i$ would be sections instead of functions.} with respect to the explicit $\vartheta$-dependence of $\mu$. The deformed supersymmetry algebra again closes to
\begin{equation}
\tilde{\delta}^2=-2\ii\mathcal{L}_v~.
\end{equation}
Note that for $\partial\mu=0$ there are no terms in the algebra mixing equivariance with the complex structure deformation. In this case, by turning off equivariance, we return to the theory considered in \cite{Labastida:1994ss}.

The deformed algebra can be expressed compactly using superfields by modifying $\delta{\bf\Phi}^i$ in \eqref{eq:sf.transf} accordingly:
\begin{equation}
  \tilde\delta{\bf\Phi}^i=\mathcal{D}_v{\bf\Phi}^i+\frac{\ii}{2}\mu^i{}_{\bar j}{\bf H}^{\bar j}+\frac{\kappa}{\|v\|^2}\wedge\partial\mu^i{}_{\bar j}{\bf H}^{\bar j}~,
\end{equation} 
where $\mu^i{}_{\bar j}(\bf\Phi,\bar\phi,\vartheta)$ here is a superfield. It is easy to see that in order for the algebra to close $\mu^i{}_{\bar j}$ needs to satisfy the KS equation \eqref{eq:KS}.

In the undeformed theory supersymmetric observables are given by insertions of holomorphic functions $f$ at the fixed points of $v$. For the deformed theory, when inserted at the north pole ($\vartheta=0$) the function $f$ needs to be holomorphic with respect to the deformed complex structure
\begin{equation}
  \partial_{\bar j} f(\phi,\bar\phi)+\frac{\ii}{2}\mu^i{}_{\bar j}\,\partial_if(\phi,\bar\phi)=0~,
\end{equation}
where the deformation $\mu$ is evaluated at $\vartheta=0$. Similarly when inserted at the south pole $f$ needs to be holomorphic with respect to the deformed complex structure at $\vartheta=\pi$~.

The equivariant $B$ model Lagrangian in the presence of small deformations $\mu(\vartheta)$ of the complex structure of $X$ is given by
\begin{equation}
  \mathcal{L}=\ell\,\tilde\delta\left(\ii g_{i\bar j}\rho^i\wedge\star\dd\phi^{\bar j}-\frac{1}{2}\Sigma^i\theta_i\right)~,
\end{equation}
where we have also introduced the large positive real parameter $\ell$ for localization. The bosonic and fermionic parts of $\mathcal{L}$ read
\begin{align}
    \mathcal{L}_\mathrm{bos}=&~\ell\left(2g_{i\bar j}\dd\phi^i\wedge\star\dd\phi^{\bar j}-\ii g_{i\bar j}{\mu^i}_{\bar k}\dd\phi^{\bar k}\wedge \star\dd\phi^{\bar j}-\frac{1}{2}\Sigma^i\beta_i\right),\\[.5em]
    \mathcal{L}_\mathrm{fer}=&~\ell\left(-\ii g_{i\bar j}\rho^i\wedge\star\dd^\Gamma\eta^{\bar j}-\dd^\Gamma\rho^i\theta_i-\frac{\ii}{4}R^i{}_{j\bar l k}\rho^j\wedge\rho^k\eta^{\bar l}\theta_i\right.\nonumber\\
    &\phantom{~t\Bigg(}-\frac{\ii}{4}\partial_k\mu^i{}_{\bar j}\,\eta^{\bar j}\Sigma^k\theta_i-\frac{\ii}{2} D_k\mu^i{}_{\bar j}\rho^k\wedge\dd\phi^{\bar j}\theta_i+\frac{1}{8}D_lD_k\mu^i{}_{\bar j}\,\eta^{\bar j}\rho^l\wedge\rho^k\theta_i\label{eq:cs.fer}\\
    &\phantom{~t\Bigg(}\left.+\frac{1}{2} g_{i\bar j}D_l\mu^i{}_{\bar k}\rho^l\eta^{\bar k}\wedge\star \dd \phi^{\bar j}+\frac{\ii}{2}{\kappa\over \|v\|^2}\wedge \partial {\mu^i}_{\bar j}\,\eta^{\bar j}\theta_i\right)~.\nonumber
\end{align}
Note that, given the structure of the supersymmetry transformations, the only part of the Lagrangian intertwining equivariance with the complex structure deformation is the last term in $\mathcal{L}_\mathrm{fer}$ with the explicit position-dependence of $\mu$. In the case of $\partial\mu^i{}_{\bar j}\equiv0$ we thus recover the Lagrangian in \cite{Labastida:1994ss}.

In order to identify the localization locus we integrate out the auxiliary fields and rewrite the bosonic part of the Lagrangian as follows:
\begin{equation}
  \begin{aligned}
    \mathcal{L}_\mathrm{bos}=&~\ell g_{i\bar k}\left(\dd \phi^i-\frac{\ii}{2}\mu^i{}_{\bar j}\dd\phi^{\bar j}\right)\wedge \star \left(\dd \phi^{\bar k}+\frac{\ii}{2}(\mu^\ast)^{\bar k}{}_l\dd\phi^l\right)\\
    &~+\ell g_{i\bar k}\left(\dd\phi^i\wedge\star\dd\phi^{\bar k}-\frac{1}{4}\mu^i{}_{\bar j}(\mu^\ast)^{\bar k}{}_l\dd\phi^{\bar j}\wedge\star\dd\phi^l\right)\\
    &~-{\ii\over 2}\,\ell g_{i\bar k}\left({\mu^i}_{\bar j}\dd\phi^{\bar j}\wedge \star \dd \phi^{\bar k}+{(\mu^*)^{\bar k}}_{j}\dd\phi^{ j}\wedge \star \dd \phi^{ i}\right).
  \end{aligned}
\end{equation}
With the following choice of reality conditions:
\begin{equation}
  (\dd\phi^i)^\ast=\dd\phi^{\bar i}~,\qquad(\mu^i{}_{\bar j})^\ast=(\mu^\ast)^{\bar i}{}_{j}~,
\end{equation}
the third line is purely imaginary and the second line is strictly positive for small but finite deformations. The localization locus is thus determined by maps $\phi$ solving the equation
\begin{equation}
  \dd \phi^{\bar i} +{\ii\over 2}{(\mu^*)^{\bar i}}_j \dd \phi^j=0~.
\end{equation} 
Contracting this equation with the complex conjugate deformation yields
\begin{equation}
  \dd \phi^i ={1\over 4}{\mu^i}_{\bar j}{(\mu^*)^{\bar j}}_{k}\dd\phi^{k}.
\end{equation}
Generically (and certainly for small enough deformations), the solutions to this equation are again constant maps $\Phi: S^2 \rightarrow X$, for which $\dd\phi^i=0$ as in the undeformed case.

For $\partial\mu=0$ the fermionic fields $\eta^{\bar i},\theta^{\bar i}$ have zero-modes $\eta^{\bar i}_0,\theta^{\bar i}_0$ and the partition function vanishes. However, when $\partial\mu\neq0$, the last term in \eqref{eq:cs.fer} lifts these zero-modes. Indeed for constant $\phi^i$ and  $\eta^{\bar i}=\eta^{\bar i}_0~, \theta^{\bar i}=\theta^{\bar i}_0$ we have:
\begin{equation}
  \int {\kappa\over \|v\|^2}\wedge \partial {\mu^i}_{\bar j}\,\eta_0^{\bar j}\,\theta_{0i}= {A^i}_{\bar j}\,\eta_0^{\bar j}\,\theta_{0i}~,
\end{equation}
where $A=R\epsilon^{-1}\mu(\vartheta)\big|_0^\pi\in\Omega^{0,1}_X(T^{1,0}X)$. Consequently, we obtain a non-vanishing partition function in this case,
\begin{equation}
\label{csref}
  \mathcal{Z}\propto\int_X \langle A^{\wedge\dim_\mathbb{C}X},\Omega_0\rangle\wedge\Omega_0~,
\end{equation}
where $\Omega_0$ denotes the holomorphic top-form on $X$ with respect to the undeformed complex structure.
The proportionality factor reflects the normalization of $\Omega_0$ and a one-loop factor which is independent
of the equivariance parameter $\epsilon$. 

Consider the special case where $\mu$ vanishes at one of the poles, say $\mu|_{\vartheta=0}=0$. We then have a smooth deformation from the original complex structure at the north pole to a deformed one $\tilde\mu:=\mu|_{\vartheta=\pi}$ at the south pole and \eqref{csref} is given by
\begin{equation}
  \int_X\Omega_{\tilde\mu}\wedge\Omega_0~,
\end{equation} 
where $\Omega_{\tilde\mu}=\Omega_0+\sum_{k=1}^{\dim_\mathbb{C}X}\langle\tilde\mu^{\wedge k},\Omega_0\rangle$ is the holomorphic top-form with respect to the deformed complex structure.

As a very simple example we can consider the equivariant $B$ model with $T^2$ target space, which is a free theory. In this case $\Omega^{0,1}_X(T^{1,0}X)$ is one dimensional and target space complex structure deformations are parametrized by a single complex parameter which we will take proportional to $\cos(\vartheta)$. Let the undeformed complex structure at $\vartheta={\pi\over 2}$  correspond to modular parameter $\tau=\ii$ so that $\phi\sim\phi+1$ and $\phi\sim \phi+\ii$. At any given $\vartheta$ a holomorphic function satisfies
\begin{equation}
  \bar \partial f+\ii t\cos(\vartheta)\partial f=0~.
\end{equation} so that it only depends on $\phi-\ii \,t\cos(\vartheta)\bar\phi$~. Hence the deformation of the complex structure is described by a $\vartheta$-dependent modular parameter $\tau$ related to $t$ by \begin{equation}
\tau(\vartheta)=\ii-{2t\cos(\vartheta) \over 1- \ii \,t\cos(\vartheta)}~.
\end{equation}
The Lagrangian is:
\begin{align}
    \mathcal{L}_\mathrm{bos}=&~2\dd\phi \wedge\star\dd\bar \phi-2\ii t \cos(\vartheta) \dd\bar \phi\wedge \star\dd\bar \phi-\frac{1}{2}\Sigma^i\beta_i~,\nonumber\\[.5em]
    \mathcal{L}_\mathrm{fer}=&-\ii \rho \wedge\star\dd\eta-\dd\rho\,\theta +\frac{t}{R \epsilon}\eta \,\theta\star 1~.
\end{align}
Excluding the constant modes for $\eta,\,\theta$ and $\phi$ the one-loop determinants are as in the undeformed case and cancel between bosons and fermions. The constant modes for $\eta,\,\theta$ are lifted by the deformation and are not zero modes as in the undeformed theory. As a result the partition function is linear~in~$t$ in accord with~\eqref{csref}. 

\section{Summary and outlook}\label{sec:summary}

In this work we have explored the equivariant deformation of the $B$ model on $S^2$. In particular we concentrated on the study of a new class of observables and their correlators. 
The equivariant $B$ model has some curious features. An interesting point is that away from the fixed points $v\neq 0$ the equivariant $B$ model transformations (\ref{transfb}) can be mapped to the equivariant $\bar{B}$ model. Thus it appears that the equivariant extension of the $B$ model stands out compared to other equivariant extensions of real models. 
   
 We defined the equivariant Landau-Ginzburg model with a
 superpotential $W(\phi, \vartheta)$ with explicit worldsheet dependence. We argued that the model localizes on tracking solutions which relate critical points of $W$ at two poles of $S^2$. We argued that these tracking solutions exist but it is hard to analyse them explicitly and perform a one-loop calculation around them. It remains to be seen if the proposed structures lead to any new interesting mathematical construction. We hope to come back to these issues in the future.  

 In the context of this paper we want to point out that one could glue together the equivariant $B$ model on one hemisphere of $S^2$ with the equivariant $\bar{B}$ model defined on another hemisphere in the spirit of~\cite{Dedushenko:2018aox,Dedushenko:2018tgx}. This model would then correspond to the theories studied in
 \cite{Doroud:2013pka}. We plan to write a separate work \cite{B-barB} on this equivariant $B/\bar{B}$ model on $S^2$ using some of the elements of the formalism presented here. 

\subsubsection*{Acknowledgements}

We are grateful to Chiara Carpenter-Festuccia for lively discussions during this project. 
 The research of G.\ F. is supported by the VR project grant 2024-05059.
 R.\ M. acknowledges support from the Centre for Interdisciplinary Mathematics at Uppsala University. The research of M.\ Z.\ 
   is  supported by the VR excellence center grant ``Geometry and Physics'' 2022-06593.

\appendix

\section{Relation to \texorpdfstring{${\cal N}=(2,2)$}{N=(2,2)} theory on squashed \texorpdfstring{$2$-sphere}{2-sphere}}
\label{app:NLSM}

Here we first review how to place a ${\cal N}=(2,2)$ theory on a squashed two sphere preserving two supercharges that give rise to an equivariant deformation of the topological twist. We follow the analysis in~\cite{Closset:2014pda} to which we refer the reader for further details. In particular we consider a nonlinear sigma model with K\"ahler target space parametrized by twisted chiral multiplets. By rewriting the twisted chiral fields in terms of cohomological variables we will make contact with the models considered in the paper.
 
\subsection{Killing spinors}
 
In order to preserve supersymmetry in curved space we couple a theory to background supergravity. For the case of a ${\cal N}=(2,2)$ theory the supergravity multiplet includes several bosonic fields in addition to the metric~\cite{Closset:2014pda}. These are  two auxiliary scalars $\cal H$ and $\widetilde{\cal H}$ and a connection $a^{(R)}$ for the $U(1)$ R-symmetry. Setting to zero the gravitino variation we obtain the Killing spinor equations
 \bea\label{Killing}
 &&(\nabla_m -\ii a^{(R)}_m) \zeta = -\frac{1}{4} {\cal H}\gamma_m (1-\gamma^3) \zeta -\frac{1}{4} {\widetilde{\cal H}}  \gamma_m(1+\gamma^3) \zeta~,\\[.5em]
 &&(\nabla_m +\ii a^{(R)}_m) \tilde \zeta = -\frac{1}{4} {\cal H} \gamma_m (1+\gamma^3)\tilde \zeta -\frac{1}{2} {\widetilde {\cal H}}\gamma_m (1-\gamma^3)\tilde \zeta~.
 \eea
Each solution to these equations corresponds to a supercharge acting via supersymmetry variations $\delta_\zeta$ and $\delta_{\tilde \zeta}$. For two solutions $\zeta$ and $\tilde \zeta$ we can define the spinor bilinears:
 \be
 \label{bildef}
 v^m= -2\zeta\gamma^m\tilde \zeta~,\qquad s= \tilde \zeta(1-\gamma_3)\zeta~,\qquad \tilde s= \tilde \zeta(1+\gamma_3)\zeta~.
 \ee
On any field\footnote{We set the central charges of $\phi$ to zero. The case with nonzero central charges is considered in~\cite{Closset:2014pda}} $\phi$ of R-charge $r$ the algebra satisfied by the variations $\delta_\zeta$ and $\delta_{\tilde \zeta}$ is
 \bea
 \label{algsimp}
 && \{\delta_\zeta, \delta_{\tilde \zeta}\} \phi = \ii ({\cal L}_v-\ii r v^m a^{(R)}_m)\phi- \frac{i}{2} r s {\cal H} \phi- \frac{i}{2} r \tilde s \widetilde {\cal H} \phi~,\\[.5em]
 && \{\delta_\zeta, \delta_{\zeta}\} \phi=0,\qquad \{\delta_{\tilde \zeta}, \delta_{\tilde \zeta}\} \phi=0~.
 \eea
As in~\eqref{metsq} consider the $2$-sphere with metric
$$
ds^2=R^2 f(\vartheta)^2(d\vartheta^2+\sin(\vartheta)^2 d\varphi^2)~.
$$
Away from the pole at $\vartheta=\pi$ it is convenient to introduce complex coordinates 
\bea
z=\tan(\vartheta/2)\e{i\varphi}~,\quad \bar z =\tan(\vartheta/2)\e{-i\varphi}~,
\eea
in terms of which the metric reads
\bea
ds^2={4 c^2( z \bar z) R^2\over (1+ z \bar z)^2} dz  d\bar z~,\qquad c(\tan^2(\vartheta/2))=f(\vartheta)~.
\eea
With the choice of the orthonormal frame:
 \be\label{roundmet}
 ds^2= \e{1} \e{\bar 1}~,\qquad \e{1}= \frac{2 R c(z \bar z)}{1+z \bar z}dz~.
 \ee
we define the spinors 
 \begin{equation}
 \label{2dspin}
 \begin{aligned}
 &\zeta_*=\begin{pmatrix}
    \zeta_-  \\ \zeta_+
  \end{pmatrix}={1\over \sqrt{2}}\begin{pmatrix}
  -\ii \epsilon {\bar z  \over 1+z \bar z}  \\
 1
  \end{pmatrix}~, \\[.5em]
 & \tilde \zeta_*=\begin{pmatrix}
   \tilde \zeta_-  \\ \tilde \zeta_+
  \end{pmatrix}=\frac{1}{\sqrt{2}}\begin{pmatrix}
 - 1  \\
   \ii \epsilon { \bar z  \over 1+z \bar z} 
  \end{pmatrix}~.
\end{aligned}
\end{equation}
 
The spinors in~\eqref{2dspin} are solutions of the Killing spinor equations~\eqref{Killing} in a patch around the origin.  The two scalars $\cal H$ and $\widetilde{\cal H}$ are given by
 \be
 {\cal H}=-\ii {\epsilon\over 2 R}\left({1- z \bar z\over 1+z \bar z}+2 z \bar z {c'\over c} \right)~,\qquad  \tilde {\cal H}= 0~.
 \ee  
The background $U(1)_R$ connection is the same as for the topological twist. It has unit flux through the $S^2$, and can locally be written as:
\bea
a^{(R)}={\ii\over2}\left({ \bar z  \over 1+ z \bar z}-{c'\over c}\right)dz-{\ii\over2}\left({ z  \over 1+ z \bar z}-{c'\over c}\right)dz~.
\eea
The spinors bilinears~\eqref{bildef}, built from $\zeta$ and $\tilde \zeta$ are then 
\be
 v= {\ii\epsilon\over R}(z\partial_z -\bar z \partial_{\bar z })~,\quad s=1~,\quad \tilde s =\epsilon^2 c^2 {z \bar z\over (1+z \bar z)^2}~.
\ee
Thus we have an equivariant deformation of the topological twisted supercharges. In terms of background supergravity this necessitates turning on the scalar $ {\cal H}$ in addition to the background $U(1)_R$ connection.

\subsection{Twisted chiral multiplet}
\label{twichi}
  
In order to describe the equivariant $B$ model we will couple twisted chiral multiplets to the supergravity background described above. Following~\cite{Closset:2014pda} and \cite{Closset:2015rna} we will review the supersymmetry transformations and actions for the twisted chiral multiplets and finally rewrite the multiplet in terms of the cohomological variables we use. 
A twisted chiral multiplet has components $\big(\phi^i, \eta^i_- , \tilde \eta^i_+, G^i\big)~,$
of $R$-charges $0, 1, -1,0$, respectively. Their supersymmetry variations are 
\bea\label{TdChl}
\delta \phi^i &=& \sqrt{2} (\tilde\zeta_+\eta^i_- - \zeta_-\tilde\eta^i_+)~, \cr
\delta \eta^i_- &=& {1\over \sqrt{2}}(1+\gamma_3)\left(\zeta G^i -\ii \gamma^m \zeta \partial_m \phi^i \right)~, \cr
\delta \tilde\eta^i_+ &=&  {1\over \sqrt{2}}(1-\gamma_3)( \tilde\zeta_+ G^i -\ii \gamma^m \tilde\zeta \partial_m \phi^i )~, \cr
\delta G^i &=& -{\ii \over \sqrt{2}}  (\zeta (1+\gamma_3)\gamma^m D_m \tilde\eta^i + \tilde\zeta (1-\gamma_3)\gamma^m D_m \eta^i)\,.
\eea
The twisted antichiral multiplet similarly has components
$
 \big(\phi^{\bar i}, \tilde\eta^{\bar i}_- , \eta^{\bar i}_+,  G^{\bar i}\big)~,
$
of $R$-charges $0, -1, 1, 0$, respectively. Their  supersymmetry transformations are 
\bea\label{TwdAChi}
\delta \phi^{\bar i} &=& -\sqrt{2} (\zeta_+\tilde\eta^{\bar i}_- - \tilde\zeta_-\eta^{\bar i}_+)~, \cr
\delta \tilde\eta^{\bar i}_- &=& {1\over \sqrt{2}}(1+\gamma_3)(\tilde\zeta  G^{\bar i} +\ii\gamma^m \tilde\zeta \partial_m\phi^{\bar i})~, \cr
\delta \eta^{\bar i}_+ &=& {1\over \sqrt{2}}(1-\gamma_3)( \zeta  G^{\bar i} +\ii  \gamma^m \zeta \partial_m \phi^{\bar i})~, \cr
\delta  G^{\bar i} &=& -{\ii\over  \sqrt{2}}  (\tilde\zeta (1+\gamma_3)\gamma^m D_m \eta^{\bar i} + \zeta(1-\gamma_3) \gamma^m D_{m} \tilde\eta^{\bar i})\,.
\eea
The sigma model action for twisted chiral fields parametrizing a K\"ahler target space with superpotential $W(\phi^i)~,\widetilde W(\phi^{\bar i})$ is as follows:
\begin{equation}
\label{dtwistedchiral}
\begin{aligned}
{\cal L} =~& g_{i\b j}\, \Big( \partial_m \phi^i \partial^m \phi^{\b j}- (G^i+\Gamma^i_{l k} \eta_-^l\tilde \eta_+^k ) (G^{\bar j}+\Gamma^{\bar j}_{\bar n \bar m} \eta_+^{\bar n}\tilde \eta_-^{\bar m})\Big)\\
& -{\ii\over 2} g_{i\bar j} \left(\eta^{\bar j} (1+\gamma_3)\gamma^m { D^\Gamma}_m \tilde\eta^i + \tilde\eta^{\bar j}(1-\gamma_3)\gamma^m { D^\Gamma}_{m} \eta^i \right) +R_{i\bar k, j \bar n}\, \tilde\eta_+^i \tilde\eta_-^{\bar k} \eta_+^{\bar n} \eta_-^j  \\
& +(G^i+\Gamma^i_{j k} \eta_-^j\tilde \eta_+^k ) \partial_i W+\eta_-^j \tilde \eta_+^k D_j\partial_k W+i\tilde {\cal H} W\\
&+(G^{\bar i}+\Gamma^{\bar i}_{\bar j \bar k} \eta_+^{\bar j}\tilde \eta_-^{\bar k})\partial_{\bar i} \widetilde W+\eta_+^{\bar j}\tilde \eta_-^{\bar k}  D_{\bar j}\partial_{\bar k} \widetilde W-i {\cal H} \widetilde W~.
\end{aligned}
\end{equation}
The coupling to the background supergravity enters through the covariant derivatives, which include the coupling to the background $U(1)_R$ connection, and through ${\cal H}$ and $\tilde {\cal H}$. We also define 
\bea
&&  { D^\Gamma}_{m} \tilde\eta^i =  D_m \tilde\eta^i + \Gamma^i_{jk} (\partial_m \phi^k) \tilde\eta^j~,\cr
&& { D^\Gamma}_{m} \eta^i =  D_m \eta^i + \Gamma^i_{jk} (\partial_m \phi^k) \eta^j~,\cr
&& D_j \partial_k W=\partial_j\partial_k W-\Gamma^i_{j k}\partial_i W~,\cr
&& D_{\bar j} \partial_{\bar k} \widetilde W=\partial_{\bar j}\partial_{\bar k} \widetilde W-\Gamma^{\bar i}_{\bar j \bar k}\partial_{\bar i}\widetilde W~.
\eea

We want to rewrite the twisted chiral multiplet in terms of cohomological variables of R-charge 0. 
\begin{align}
&&\phi^i,\quad \rho^i_m={1\over\sqrt{2} s}\big(\tilde \zeta(1\!-\!\gamma_3)\gamma_m \eta^i- \zeta (1\!+\!\gamma_3)\gamma_m \tilde \eta^i\big), \quad \Sigma^i_{m n}= {1\over s} (G^i \!+\Gamma^i_{j k} \eta_-^j \tilde \eta_+^k)\epsilon_{m n} ~,\nonumber \\
&&\phi^{\bar i},\quad  \eta^{\bar i}=\delta\phi^{\bar i},\quad \theta^{\bar i}={1\over \sqrt{2}}(\tilde \zeta(1- \gamma_3 )\eta^i-\zeta (1+\gamma_3)\tilde \eta^i),\quad \beta^{\bar i}=2 s (G^{\bar i}\!+\Gamma^{\bar i}_{\bar j \bar k} \eta^{\bar k}_+\tilde \eta_-^{\bar j} ) ~.
\end{align} 
An equivalent set of variables is described in~\cite{Closset:2015rna}~.
The map is well-defined and invertible because on our background the spinor bilinear $s$ is a nonzero constant. By using the transformations~\eqref{TdChl} and~\eqref{TwdAChi} we can find the action of supersymmetry on the twisted variables reproducing~\eqref{transfb}.

The definitions of $\rho^i, \eta^{\bar i}, \theta^{\bar i}$ and $s$ only involve the Killing spinor components $\zeta_+$ and $\tilde\zeta_-$. From~\eqref{2dspin} these are constant and independent of $\epsilon$. In particular this implies that the definition of the twisted variables is the same with or without equivariance. As a simple consequence the expression for the Lagrangian~\eqref{dtwistedchiral} in the twisted variables is the same with or without equivariance apart from the term involving ${\cal H}$ which is turned on only in the equivariant setting.

\section{Second supercharge}
\label{app:susy}
In this appendix we list the field transformations corresponding to the second supercharge of the $B$ model and its equivariant deformation (see also~\cite{Closset:2015rna}). They are given by
\begin{equation}
  \begin{aligned}\label{transfbtwoa}
\hat\delta \phi^i &=\ii\,\iota_v \star\!\rho^i~,\qquad \hat\delta \rho^i=-2\star\!\dd \phi^i+\kappa \star(\Sigma^i-{\ii\over 2}\Gamma^i_{j k}\rho^j\wedge\rho^k)~,\\
\hat\delta\Sigma^i &=2\ii \dd^\Gamma\! \star \rho^i+\ii \Gamma^{i}_{j k} \star\! \Sigma^j \kappa\wedge \rho^\kappa+\frac{\ii}{2}R^i{}_{j\bar{l}k}\theta^{\bar l}\rho^j\wedge\rho^k~,\\
\hat \delta \phi^{\bar i}&=\theta^{\bar i},\qquad \hat \delta \theta^{\bar i}=2 \ii \iota_v \dd\phi^{\bar i},\qquad \hat \delta \eta^{\bar i}=-\beta^{\bar i}-2\star\! (\kappa\wedge d\phi^{\bar i})-\Gamma^{\bar i}_{\bar{j}\bar{k}}\theta^{\bar j}\eta^{\bar k}~,\\
\hat\delta\beta^{\bar i}&=-2\ii\iota_v\dd^\Gamma\eta^{\bar i}-2\star(\kappa\wedge\dd^\Gamma\theta^{\bar i})-\Gamma^{\bar i}_{\bar{j}\bar{k}}\theta^{\bar j}\beta^{\bar k}-\ii R^{\bar i}_{\bar{j}l\bar{k}}\iota_v\star\!\rho^l\theta^{\bar j}\eta^{\bar k}.
\end{aligned}
\end{equation}
In the non-equivariant limit the transformations above reduce to those for the standard $B$ model:

\begin{equation}
  \label{transfbtwob}
  \begin{aligned}
    &\hat\delta \phi^i=0~,\quad \hat\delta \rho^i=-2\star\!\dd \phi^i~,\qquad  \hat\delta\Sigma^i=2i \dd^\Gamma\! \star \rho^i+\frac{\ii}{2}R^i{}_{j\bar{l}k}\theta^{\bar l}\rho^j\wedge\rho^k~,
    \cr  &\hat \delta \phi^{\bar i}=\theta^{\bar i}~,\quad \hat \delta \theta^{\bar i}=0~,\quad \hat \delta \eta^{\bar i}=-\beta^{\bar i}-\Gamma^{\bar i}_{\bar{j}\bar{k}}\theta^{\bar j}\eta^{\bar k}~,\quad \hat\delta\beta^{\bar i}=-\Gamma^{\bar i}_{\bar{j}\bar{k}}\theta^{\bar j}\beta^{\bar k}~,
  \end{aligned}
\end{equation}
As was done for the first supercharge in the main text, also here we can use $\beta_i$ and $\eta_i=g_{i \bar j}\eta^{\bar j}$ in terms of which
\begin{align}
    \hat\delta \eta_i= \beta_i~,\qquad \hat \delta \beta_i= 0~.    
\end{align}

The $B$ model Lagrangian $\mathcal{L}_\mathrm{D}$ is $\hat\delta$-exact, indeed $\mathcal{L}_\mathrm{D}=\hat\delta\hat{\mathcal{V}}$ where
\begin{equation}
\label{lagdeltahat}
\hat{\mathcal{V}}=g_{i\bar j}\rho^i\wedge\dd\phi^{\bar j}+\frac{1}{2}\Sigma^i\eta_i~.
\end{equation}
Finally the observables \eqref{eq--O} satisfy descent relations with respect to $\hat\delta$:
\begin{equation}
  \hat \delta\mathcal{O}_{0}=\ii \iota_v\star\!\mathcal{O}_1~,\qquad \hat\delta\star\!\mathcal{O}_1=\ii \iota_v\mathcal{O}_2+2\dd\mathcal{O}_0~,\qquad \hat \delta\mathcal{O}_2=2\dd\mathcal{\star O}_1~.
\end{equation}

\section{Localization for \texorpdfstring{$W(\phi,\vartheta)$}{}}
\label{Apploc}
In this appendix we consider localization in the presence of the more general observables discussed in~\ref{genob}. These observables are built from an analytic function of the $\phi^i$ that also has explicit $\vartheta$-dependence $W(\phi,\vartheta)$. 

~

\noindent Up to a total derivative the bosonic Lagrangian we use for localization is 
\begin{equation}
\begin{aligned}
    \mathcal{L}_\mathrm{bos}=&~(1+\ell||v||^2)\left(2g_{i\bar j}\dd\phi^i\wedge\star\dd\phi^{\bar j}-\frac{1}{2}\Sigma^i\beta_i\right)\\
    &+{\ifthenelse{\boolean{eps}}{\epsilon}{1}\over 2}(1+\ell ||v||^2)(\partial_i W)\Sigma^i+{\ifthenelse{\boolean{eps}}{\epsilon}{1}\over 2} \ell(\partial_{\bar i}\widetilde{W})\beta^{\bar i}\star 1\\
    &-{\ifthenelse{\boolean{eps}}{\epsilon}{}\kappa\over ||v||^2}\wedge \partial W+\ell\ifthenelse{\boolean{eps}}{\epsilon}{} \kappa\wedge (\partial_{\bar i}{\widetilde W}d\phi^{\bar i})+\ell\ifthenelse{\boolean{eps}}{\epsilon}{} \kappa\wedge( \partial_i W d\phi^i)~.
\end{aligned}
\end{equation}
The auxiliary fields can be integrated out
\begin{equation}
\label{auxintb}
    \Sigma^i=\frac{\ell\ifthenelse{\boolean{eps}}{\epsilon}{}}{1+\ell||v||^2}g^{i \bar j}\partial_{\bar j}\widetilde{W} \star 1~,\qquad \beta^{\bar i}=\ifthenelse{\boolean{eps}}{\epsilon}{}g^{j \bar i}\partial_{ j}{W}~.
\end{equation}
which results in the bosonic Lagrangian
\begin{equation}
    \begin{aligned}
        \mathcal{L}_\mathrm{bos}=&~2(1+\ell ||v||^2)g_{i\bar j}\dd\phi^i\wedge\star\dd\phi^{\bar j}+\frac{\ell\ifthenelse{\boolean{eps}}{\epsilon}{}}{2}g^{i\bar j}\left(\partial_{ i}{W}\right)\wedge\star\left(\partial_{\bar j}\widetilde{W}\right)\\
          &-{\ifthenelse{\boolean{eps}}{\epsilon}{}\kappa\over ||v||^2}\wedge \partial W+\ell\ifthenelse{\boolean{eps}}{\epsilon}{} \kappa\wedge (\partial_{\bar i}{\widetilde W}d\phi^{\bar i})+\ell\ifthenelse{\boolean{eps}}{\epsilon}{} \kappa\wedge( \partial_i W d\phi^i)~.
     \end{aligned}
\end{equation}
completing squares yields
 \begin{equation}
 \label{squaregen}
    \begin{aligned}
        \mathcal{L}_\mathrm{bos}=~&2 g_{i\bar j}\dd\phi^i\wedge\star\dd\phi^{\bar j}+2\ell g_{i\bar j}\iota_v \dd \phi^i \iota_v\dd \phi^{\bar j} \star 1  -{\ifthenelse{\boolean{eps}}{\epsilon}{}\kappa\over ||v||^2}\wedge \partial W \\
        &+2\ell g_{i \bar j}\left(\star(\kappa \wedge d\phi^{\bar j})+{\ifthenelse{\boolean{eps}}{\epsilon}{1} \over 2}\partial^{\bar j}W\right)\left(\kappa\wedge d\phi^{i}+{\ifthenelse{\boolean{eps}}{\epsilon}{1}\over 2}\partial^{i} W\star 1\right).     
        \end{aligned}
\end{equation}
Hence the functional integral localizes to scalar field configurations that satisfy 
\begin{equation}
\label{eqtrkuu}
\iota_v \dd \phi^{\bar i}=0~,\qquad \star (\kappa\wedge \dd \phi^{\bar i})+{\ifthenelse{\boolean{eps}}{\epsilon}{1}\over 2}\,g^{j\bar i} \left(\partial_{ j}{W}\right)=0~.
\end{equation}
As for the case considered in the main text these two equations are the BPS conditions $\delta\eta^{\bar i}=0$ and $\delta \theta^{\bar i}=0$. The remaining BPS condition $\delta \rho^i=0$ is satisfied on solutions to~\eqref{eqtrkuu} that are regular at the poles.
Finally let's consider evaluating the bosonic action on the localization locus. Because of~\eqref{eqtrkuu} we have:
\begin{equation}
    \star d\phi^{\bar i}={\ifthenelse{\boolean{eps}}{\epsilon}{}\kappa\over 2||v^2||}g^{j \bar i}\partial_j W~.
\end{equation}
We can then substitute this back into~\eqref{squaregen} to get:
\begin{equation}
     \mathcal{L}_\mathrm{bos}= (\partial_i W d\phi^i+\partial W)\wedge{\ifthenelse{\boolean{eps}}{\epsilon}{}\kappa\over ||v||^2}= dW\wedge {\ifthenelse{\boolean{eps}}{\epsilon}{}\kappa\over ||v||^2}~,
\end{equation}
so that
\begin{equation}
    S=-\ifthenelse{\boolean{eps}}{}{{1\over \epsilon}}\int_{S^2} d\varphi \wedge dW=-\ifthenelse{\boolean{eps}}{2\pi}{{2\pi\over \epsilon}} \int_0^\pi {d\over d\vartheta}W d\vartheta= \ifthenelse{\boolean{eps}}{2\pi}{{2\pi\over \epsilon}} (W|_{\vartheta=0}-W|_{\vartheta=\pi})~.
\end{equation}

\section{Domain walls}
\label{Appdom}
In this appendix we consider the special case of the minimal model $W_1(\phi)=\frac{1}{3}\phi^3-\phi$ with $W_2\equiv0$. Here, the BPS equation \eqref{eqtrb} allows for domain wall solutions that interpolate between different critical points of $W_1$. However, we will see that these solutions come with corresponding fermionic zero-modes so that their contribution to the correlators we are interested in vanishes.

~

\noindent For simplicity, we consider the round sphere with $R=1$ and $\epsilon=1$, but the same analysis goes through for a generic choice of background. Upon substituting $\binoppenalty=10000 
\relpenalty=10000  x=\frac{1}{2}\log(\tan(\vartheta/2))$ into \eqref{eqtrb} for the choice of $W_1,W_2$ above, we obtain the following equation:
\begin{equation}\label{eq:dom.wall}
    \partial_x\phi-\partial \widetilde W_1=0~.
\end{equation}
Note that $\phi=\phi(x)$ does not depend on the azimuthal angle because of the BPS condition $\iota_v\phi=0$. The critical points of the potential are $\phi\equiv\pm1$ and the domain wall solution interpolating between them is real and given by $\phi_0(x)=\tanh(x-a)$. Here, $a\in\mathbb{R}$ denotes the center of the domain wall which is a bosonic zero-mode.

In order to look for fermionic zero-modes we need to study the Lagrangian
\begin{equation}
\label{lafr}
    \mathcal{L}_\mathrm{fer}=(1+\ell\|v\|^2)(-\ii\rho\wedge\star\dd\bar\eta-\dd\rho\bar\theta+\frac{\ii}{4}\partial^2W(\phi_0)\rho\wedge\rho)+\frac{\ell}{2}\partial^2\widetilde{W}(\bar\phi_0)\bar\eta\bar\theta\star1~,
\end{equation}
as the corresponding fermionic part to \eqref{eq:Lfull}. Zero-modes of this Lagrangian have to satisfy the following differential equations:
\begin{subequations}\label{eq:zm}
    \begin{align}
        0&=\frac{\ii}{4}(1+\ell\|v\|^2)\partial^2W(\phi_0)\star\rho-\frac{1}{2}\star\dd((1+\ell\|v\|^2)\bar\theta)+\frac{\ii}{2}(1+\ell\|v\|^2)\dd\bar\eta~,\label{eq--wall.zm.a}\\
        0&=(1+\ell\|v\|^2)\star\dd\rho-\frac{1}{2}\ell\partial^2\widetilde{W}(\bar\phi_0)\bar\eta~,\label{eq--wall.zm.b}\\
        0&=\ii\dd^\dagger((1+\ell\|v\|^2)\rho)+\frac{\ell}{2}\partial^2\widetilde{W}(\bar\phi_0)\bar\theta~.\label{eq--wall.zm.c}
    \end{align}
\end{subequations}
Zero modes are $v$ invariant hence we impose $\mathcal{L}_v\bar\theta=\mathcal{L}_v\bar\eta=0$ and $\mathcal{L}_v\rho=0$

We will start by finding the zero modes for the terms of order $\ell$ in~\eqref{lafr}. Decomposing $\rho=\rho_{||}+\rho_\perp$ where $\iota_v \rho_\perp=0$ and $\kappa \wedge \rho_{||}=0$ we get the uncoupled equations
\begin{align}
& d^\dagger (\phi_0^{-1} d\bar\eta)+{1\over||v||^2}\phi_0 \bar\eta=0~, \\
& d^\dagger (\phi_0^{-1} d(||v^2||\bar\theta))+{1\over ||v||^2}\phi_0 (||v||^2 \bar\theta)=0~.
\end{align}
These two equations have the same form and are easily solved by:
\begin{align}
&\bar \eta= c_1 \sech(x-a)^2+c_2 \cosh(x-a)^2~,\\
&\bar \theta= c_3 \sech(x-a)^2\cosh(2x)^2 +c_4 \cosh(x-a)^2 \cosh(2x)^2~.
\end{align}
Only one solution is normalizable hence there is one zero mode:
\begin{equation}
    \rho=c\sech^2(x-a)\frac{\kappa}{\|v\|^2}~,\qquad\bar\eta=c\sech^2(x-a)~,\qquad\bar\theta\equiv0~.
\end{equation}
Here, $c\in\mathbb{R}$ is a constant and is the fermionic superpartner of $a$, $\delta a=c$. Hence, the existence of the fermionic zero-mode is linked to the bosonic one. This will be the case also for other potentials $W_1$ allowing for domain wall solutions. Note that the fermionic terms in the localization Lagrangian~\eqref{lafr} explicitly break $\hat \delta$. Hence there is no fermionic zero mode related to the bosonic one by the second supercharge $\hat \delta$.

We can now consider all terms in~\eqref{lafr}. As zero modes can be lifted only in pairs and the terms of order $\ell$ have only one zero mode this will survive the addition of the terms of order $\ell^0$.

\section{Chiral ring relations}
\label{app:chiral}

As we discussed, in the equivariant $B$ model, a holomorphic function of the $\phi^i$ inserted at a fixed point, say the north pole at $\vartheta=0$, is supersymmetric
\begin{equation}
    {\cal O}=F(\phi)|_{N}~,\qquad \delta {\cal O}=0~.
\end{equation}
In the presence of a superpotential $W(\phi,\vartheta)$, that can depend explicitly on $\vartheta$, there are relations between supersymmetric insertions. Indeed, consider 
\begin{equation}
    \hat {\cal O}= \hat F^i(\phi)\beta_i~,
\end{equation}
where the $F^i(\phi)$ are holomorphic. 
We can write
\begin{equation}
   \delta (\hat F^i\theta_i)- \hat {\cal O}=\partial_j \hat F^i\,\iota_v \rho^j\,\theta^i+2 g_{i\bar j}\hat F^i \star (\kappa\wedge \dd \phi^{\bar j})~.
\end{equation}
The two operators on the right hand side, which are proportional to $v$ should be defined so that they vanish when inserted at the two poles. Then $\hat {\cal O}$ inserted at the fixed points of $v$ is $\delta$-exact.
Using the equations of motion we have $\beta_i=\partial_i W$ hence 
\begin{equation}
    \hat F^i\partial_i W|_{N,S}~,
\end{equation}
is $\delta$-exact on-shell.
We thus find that at the north pole $\vartheta=0$ there are relations between the $F(\phi)$ given by $\partial_i W(\phi,0)$ while at the south pole $\vartheta=\pi$ the relations are given by $\partial_i W(\phi,\pi)$.
   
\providecommand{\href}[2]{#2}\begingroup\raggedright


\providecommand{\href}[2]{#2}\begingroup\raggedright\endgroup

\endgroup


\begin{thebibliography}{10}

\bibitem{Witten:1991zz}
E.~Witten, ``{Mirror manifolds and topological field theory},'' {\em AMS/IP
  Stud. Adv. Math.} {\bfseries 9} (1998) 121--160,
  \href{http://arxiv.org/abs/hep-th/9112056}{{\ttfamily arXiv:hep-th/9112056}}.

\bibitem{Labastida:1991qq}
J.~M.~F. Labastida and P.~M. Llatas, ``{Topological matter in
  two-dimensions},'' \href{http://dx.doi.org/10.1016/0550-3213(92)90596-4}{{\em
  Nucl. Phys. B} {\bfseries 379} (1992) 220--258},
  \href{http://arxiv.org/abs/hep-th/9112051}{{\ttfamily arXiv:hep-th/9112051}}.

\bibitem{Hori:2003ic}
K.~Hori, S.~Katz, A.~Klemm, R.~Pandharipande, R.~Thomas, C.~Vafa, R.~Vakil, and
  E.~Zaslow, {\em {Mirror symmetry}}, vol.~1 of {\em Clay mathematics
  monographs}.
\newblock AMS, Providence, USA, 2003.

\bibitem{Yagi:2014toa}
J.~Yagi, ``{$\Omega$-deformation and quantization},''
  \href{http://dx.doi.org/10.1007/JHEP08(2014)112}{{\em JHEP} {\bfseries 08}
  (2014) 112}, \href{http://arxiv.org/abs/1405.6714}{{\ttfamily arXiv:1405.6714
  [hep-th]}}.

\bibitem{Nekrasov:2018pqq}
N.~Nekrasov, {\em {Tying up instantons with anti-instantons.}}
\newblock 2018.
\newblock \href{http://arxiv.org/abs/1802.04202}{{\ttfamily arXiv:1802.04202
  [hep-th]}}.

\bibitem{Closset:2014pda}
C.~Closset and S.~Cremonesi, ``{Comments on $\mathcal{N} $ = (2, 2)
  supersymmetry on two-manifolds},''
  \href{http://dx.doi.org/10.1007/JHEP07(2014)075}{{\em JHEP} {\bfseries 07}
  (2014) 075}, \href{http://arxiv.org/abs/1404.2636}{{\ttfamily arXiv:1404.2636
  [hep-th]}}.

\bibitem{Closset:2015rna}
C.~Closset, S.~Cremonesi, and D.~S. Park, ``{The equivariant A-twist and gauged
  linear sigma models on the two-sphere},''
  \href{http://dx.doi.org/10.1007/JHEP06(2015)076}{{\em JHEP} {\bfseries 06}
  (2015) 076}, \href{http://arxiv.org/abs/1504.06308}{{\ttfamily
  arXiv:1504.06308 [hep-th]}}.

\bibitem{Vafa:1990mu}
C.~Vafa, ``{Topological Landau-Ginzburg models},''
  \href{http://dx.doi.org/10.1142/S0217732391000324}{{\em Mod. Phys. Lett. A}
  {\bfseries 6} (1991) 337--346}.

\bibitem{Cecotti:1991me}
S.~Cecotti and C.~Vafa, ``{Topological antitopological fusion},''
  \href{http://dx.doi.org/10.1016/0550-3213(91)90021-O}{{\em Nucl. Phys. B}
  {\bfseries 367} (1991) 359--461}.

\bibitem{Witten:1988xj}
E.~Witten, ``{Topological Sigma Models},''
  \href{http://dx.doi.org/10.1007/BF01466725}{{\em Commun. Math. Phys.}
  {\bfseries 118} (1988) 411}.

\bibitem{Labastida:1994ss}
J.~M.~F. Labastida and M.~Marino, ``{Type B topological matter, Kodaira-Spencer
  theory, and mirror symmetry},''
  \href{http://dx.doi.org/10.1016/0370-2693(94)90158-9}{{\em Phys. Lett. B}
  {\bfseries 333} (1994) 386--395},
  \href{http://arxiv.org/abs/hep-th/9405151}{{\ttfamily arXiv:hep-th/9405151}}.

\bibitem{Cecotti:1992rm}
S.~Cecotti and C.~Vafa, ``{On classification of N=2 supersymmetric theories},''
  \href{http://dx.doi.org/10.1007/BF02096804}{{\em Commun. Math. Phys.}
  {\bfseries 158} (1993) 569--644},
  \href{http://arxiv.org/abs/hep-th/9211097}{{\ttfamily arXiv:hep-th/9211097}}.

\bibitem{Tian:1987}
G.~Tian, ``{Smoothness of the Universal Deformation Space of Compact Calabi-Yau
  Manifolds and its Peterson-Weil Metric},'' in {\em {Mathematical Aspects of
  String Theory}}, S.-T. Yau, ed.
\newblock 1987.

\bibitem{Todorov:1989umc}
A.~N. Todorov, ``{The Weil-Petersson geometry of the moduli space of
  SU(n\ensuremath{\geqq}3) (Calabi-Yau) manifolds I},''
  \href{http://dx.doi.org/10.1007/bf02125128}{{\em Commun. Math. Phys.}
  {\bfseries 126} (1989) 325--346}.

\bibitem{Dedushenko:2018aox}
M.~Dedushenko, ``{Gluing. Part I. Integrals and symmetries},''
  \href{http://dx.doi.org/10.1007/JHEP04(2020)175}{{\em JHEP} {\bfseries 04}
  (2020) 175}, \href{http://arxiv.org/abs/1807.04274}{{\ttfamily
  arXiv:1807.04274 [hep-th]}}.

\bibitem{Dedushenko:2018tgx}
M.~Dedushenko, ``{Gluing II: boundary localization and gluing formulas},''
  \href{http://dx.doi.org/10.1007/s11005-021-01355-8}{{\em Lett. Math. Phys.}
  {\bfseries 111} no.~1, (2021) 18},
  \href{http://arxiv.org/abs/1807.04278}{{\ttfamily arXiv:1807.04278
  [hep-th]}}.

\bibitem{Doroud:2013pka}
N.~Doroud and J.~Gomis, ``{Gauge theory dynamics and K\"ahler potential for
  Calabi-Yau complex moduli},''
  \href{http://dx.doi.org/10.1007/JHEP12(2013)099}{{\em JHEP} {\bfseries 12}
  (2013) 099}, \href{http://arxiv.org/abs/1309.2305}{{\ttfamily arXiv:1309.2305
  [hep-th]}}.

\bibitem{B-barB}
G.~Festuccia, R.~Mauch, and M.~Zabzine, ``{The equivariant $B/{\bar B}$
  model},'' {\em in progress} .

\end{thebibliography}
\end{document}